\documentclass[aps,prb,twocolumn,superscriptaddress,nobibnotes,amsmath,amssymb]{revtex4-2}
\usepackage{graphicx}
\usepackage{color}
\usepackage{xcolor}
\usepackage{comment}
\usepackage[explicit]{titlesec}
\usepackage[normalem]{ulem}

\titleformat{\paragraph}[runin]
	{\normalfont\small\bfseries}{}{0pt}{#1}
\titleformat{name=\paragraph,numberless}[runin]
	{\normalfont\small\bfseries}{}{0pt}{#1}
\titlespacing*{\paragraph}{0pt}{0.25cm}{0.15cm}
\titlespacing*{name=\paragraph,numberless}{0pt}{0.25cm}{0.15cm}

\begin{document} 
\newcommand*{\rom}[1]{\expandafter\@slowromancap\romannumeral #1@}
\newcommand{\change}[1]{\textcolor{blue}{#1}}
\newcommand{\remove}[1]{\textcolor{red}{\sout{#1}}}

\title{Charge-neutral nonlocal response in superconductor-InAs nanowire  hybrid devices}
\author{A.O.~Denisov}
\affiliation{Institute of Solid State Physics, Russian Academy of Sciences, 142432 Chernogolovka, Russian Federation}
\affiliation{Department of Physics, Princeton University, Princeton, New Jersey 08544, USA}
\author{A.V.~Bubis} 
\affiliation{Skolkovo Institute of Science and Technology, Nobel street 3, 121205 Moscow, Russian Federation}
\affiliation{Institute of Solid State Physics, Russian Academy of Sciences, 142432 Chernogolovka, Russian Federation}
\author{S.U.~Piatrusha}
\affiliation{Institute of Solid State Physics, Russian Academy of Sciences, 142432 Chernogolovka, Russian Federation}
\author{N.A.~Titova}
\affiliation{Moscow Pedagogical State University, 29 Malaya Pirogovskaya St, Moscow, 119435, Russia}
\author{A.G.~Nasibulin}
\affiliation{Skolkovo Institute of Science and Technology, Nobel street 3, 121205 Moscow, Russian Federation}
\affiliation{Aalto University, P. O. Box 16100, 00076 Aalto, Finland}
\author{J.~Becker}
\affiliation{Walter Schottky Institut, Physik Department, and Center for Nanotechnology and Nanomaterials, Technische Universit\"{a}t M\"{u}nchen, Am Coulombwall 4, Garching 85748, Germany}
\author{J.~Treu}
\affiliation{Walter Schottky Institut, Physik Department, and Center for Nanotechnology and Nanomaterials, Technische Universit\"{a}t M\"{u}nchen, Am Coulombwall 4, Garching 85748, Germany}
\author{D.~Ruhstorfer}
\affiliation{Walter Schottky Institut, Physik Department, and Center for Nanotechnology and Nanomaterials, Technische Universit\"{a}t M\"{u}nchen, Am Coulombwall 4, Garching 85748, Germany}
\author{G.~Koblm\"{u}ller}
\affiliation{Walter Schottky Institut, Physik Department, and Center for Nanotechnology and Nanomaterials, Technische Universit\"{a}t M\"{u}nchen, Am Coulombwall 4, Garching 85748, Germany}
\author{E.S.~Tikhonov}
\affiliation{Institute of Solid State Physics, Russian Academy of Sciences, 142432 Chernogolovka, Russian Federation}
\author{V.S.~Khrapai}
\affiliation{Institute of Solid State Physics, Russian Academy of Sciences, 142432 Chernogolovka, Russian Federation}

\begin{abstract}
Nonlocal quasiparticle transport in normal-superconductor-normal (NSN) hybrid structures probes sub-gap states in the proximity region and is especially attractive in the context of Majorana research. Conductance measurement provides only partial information about nonlocal  response composed from both electron-like and hole-like quasiparticle excitations. In this work, we show how a nonlocal shot noise measurement  delivers a missing puzzle piece in NSN InAs nanowire-based devices. We demonstrate that in a trivial superconducting phase quasiparticle response is practically charge-neutral, dominated by the heat transport component with a thermal conductance being on the order of conductance quantum. This is qualitatively explained by numerous Andreev reflections of a diffusing quasiparticle, that makes its charge completely uncertain. Consistently, strong fluctuations and sign reversal are observed in the sub-gap nonlocal conductance, including occasional Andreev rectification signals. Our results prove conductance and noise as complementary measurements to characterize  quasiparticle transport in superconducting proximity devices.
\end{abstract}

\maketitle

Nonlocal conductance measurements~\cite{PhysRevLett.77.4954} in semiconductor-superconductor proximity structures gain renewed interest in the context of Majorana  research~\cite{Lee_2017,PhysRevB.98.085125,PhysRevB.97.045421,PhysRevB.100.045302,PhysRevLett.124.036801,pan2020threeterminal,Zhao_2020}. The key underlying idea is that the nonlocal signals can probe global sub-gap states characteristic of a true topological phase transition~\cite{PhysRevLett.105.077001,PhysRevLett.105.177002,PhysRevLett.106.057001}. This is in contrast to a standard two-terminal conductance~\cite{Mourik1003,Das2012,yu2020nonmajorana} sensitive to the states near the point where the current inflows in the proximity region. Recent experiments in three-terminal NSN nanowire-based (NW-based) hybrid devices confirm conceptual power of the nonlocal conductance approach~\cite{PhysRevLett.124.036802,puglia2020closing}.

Conductance measurement provides only partial information about quasiparticle non-equilibrium in proximity structures. A sub-gap quasiparticle entering the proximity region carries the electric charge, $q<0$ for electron-like and $q>0$ for hole-like quasiparticles, and the excitation energy $\varepsilon = |E|>0$, where $E$ is the kinetic energy relative to the chemical potential of the superconductor. On its way, apart from possible normal scattering, a quasiparticle experiences a few Andreev reflections (ARs) from the superconducting lead~\cite{Andreev,Kopnin_2004}, each time inverting the $q$ but preserving the $\varepsilon$. Thereby the AR mediates a coupling of the charge and heat (energy) transport components that is unique to proximity structures and does not occur in bulk superconductors~\cite{tinkham2004introduction,Heikkila_2019}. Thus, a full characterization  of the non-equilibrium can be achieved by measurement of both the electric and heat nonlocal conductances.

In this article, we investigate the nonlocal response in NSN InAs NW-based devices. We show that a quasiparticle non-equilibrium can be understood if the nonlocal conductance is accompanied by a shot noise measurement substituting the heat conductance measurement, that allows to separate the contributions of transmission processes involving even and odd number of the ARs. Experiments performed in a trivial superconducting phase demonstrate that quasiparticle transport is practically charge-neutral, so that the heat transport component dominates the nonlocal response in our devices. Our results prove shot noise as a valuable, complementary to conductance, tool to probe the sub-gap states in proximity structures.

\begin{figure}
 	\begin{center}
 		\includegraphics[width=0.9\columnwidth]{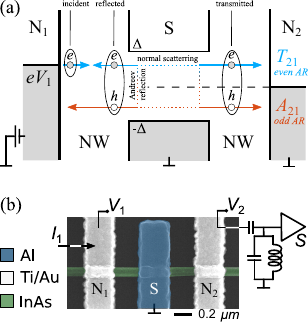}
 	\end{center}
 	\caption{(a)~Energy diagram of the NSN NW-based device. In this illustration, the left normal terminal N$_1$ is biased with voltage $V_1<0$, while the central S-terminal and the right N$_2$ terminal are grounded, their chemical potential shown by the dashed line. Depending on the number of Andreev reflections, the electron incident from N$_1$ can be transmitted towards N$_2$ as an electron ({\it e}) or hole ({\it h}) with probability $T_{\mathrm{21}}$ or $A_{\mathrm{21}}$ correspondingly. (b)~Scanning electron microscope image of the NSN-II device (false color) and the shot-noise measurement scheme used in actual experiment. Voltages $V_1$ and $V_2$ that build up in response to the current bias $I_1$ are measured in a quasi-four-terminal configuration.}
 	\label{band_dia}
 \end{figure} 

We start the discussion from the energy diagram of the NSN NW-based hybrid structure in a nonlocal experiment sketched in Fig.~\ref{band_dia}a. Consider the case of zero temperature $T=0$ and negative bias voltage $V_1<0$ applied to the left normal terminal $\mathrm{N_1}$. The superconductor and the right normal terminal $\mathrm{N_2}$ are grounded, position of their chemical potential shown by the dashed line. Transmitted quasiparticles are distinguished by their energy relative to this chemical potential, $\varepsilon>0$ for electrons and $-\varepsilon$ for holes. Inside the NW quasiparticles experience ARs from the S-lead and elastic normal scattering from disorder and possibly from the S/NW interface, inelastic scattering is absent~\cite{Tikhonov2016}. The charge current $I_\mathrm{2}$ and the heat current $J_\mathrm{2}$ in the right lead read~\cite{Claughton1996,Anantram_1996}:
%
%
%
%
	\begin{equation}
		I_\mathrm{2} = -\frac{e^2}{h} V_1\Sigma\mathcal{T}_{-};\,\,\,J_\mathrm{2} = \frac{e^2}{2h}V_1^2\Sigma\mathcal{T}_{+};\,\,\,\mathcal{T}_{\pm}\equiv T_\mathrm{21}\pm A_\mathrm{21}\label{eq1}
	\end{equation}
	%
where the positive direction for the electric current is from the lead into the scattering region and opposite for the heat current. $T_\mathrm{21}$ and $A_\mathrm{21}$  are the probabilities of transmission, respectively, preserving and changing a quasiparticle type, sometimes also called normal and crossed-Andreev transmission~\cite{PhysRevLett.124.036801} and the sum over the eigenchannels is performed. Generally, the two transmission processes involve both the normal and Andreev scattering and are distinguished by the parity of the number of ARs involved. Processes with an even and odd number of ARs contribute, respectively, to $T_\mathrm{21}$ and $A_\mathrm{21}$. Eqs.~(\ref{eq1}) imply that a simultaneous measurement of the charge and heat response permits an independent characterization of $T_\mathrm{21}$ and $A_\mathrm{21}$. Measurement of the heat transport is not an easy task~\cite{RevModPhys.78.217,Chandrasekhar_2009,Peltonen2010,PhysRevLett.117.196801} and we choose a different path in present experiment. We  perform a shot noise measurement in a nonlocal configuration~\cite{Tikhonov2016,Tikhonov_2020,Larocque_2020}, based on the findings of Ref.~\cite{PhysRevLett.106.057001} briefly mentioned below.

%
The average charge $Q_\mathrm{2}$ (in units of $e$) transmitted in one eigenchannel in an individual scattering event equals $\langle Q_\mathrm{2}\rangle = \mathcal{T}_{-}$. Its fluctuation is $\langle (\delta Q_\mathrm{2})^2\rangle = \langle Q_\mathrm{2}^2\rangle - \langle Q_\mathrm{2}\rangle^2$, where $\langle Q_\mathrm{2}^2\rangle = \mathcal{T}_{+}$. Thus, in spirit of Ref.~\cite{Landauer_1991}, we obtain for the spectral density of the current noise in the right lead:
	\begin{equation}
		S_2 = \frac{2e^3}{h}|V_1|\,\Sigma\left(\mathcal{T}_{+} - \mathcal{T}_{-}^2\right), \label{eq2} 
	\end{equation}
that contains  $\mathcal{T}_{+}$ and can substitute $J_\mathrm{2}$ in a nonlocal measurement. In the limit of suppressed AR $A_\mathrm{21}\rightarrow0$ eq.~(\ref{eq2}) reduces to a familiar result in the normal case~\cite{BLANTER20001}. In this case, a nonlocal Fano factor defined as $F_\mathrm{nl} \equiv S_2/2e|I_\mathrm{2}|$ is bounded by unity $F_\mathrm{nl}=1-T_{21}\leq1$. In the opposite limit of $T_\mathrm{21}=A_\mathrm{21}$ the shot noise and the heat current remain finite, $S_2\propto J_\mathrm{2}$, whereas $I_\mathrm{2}=0$. Here, the nonlocal quasiparticle response is charge-neutral and ${F_\mathrm{nl}\rightarrow\infty}$. 

Eqs.~(\ref{eq1}-\ref{eq2}) illustrate our main idea that nonlocal conductance $G_\mathrm{21}\equiv \partial I_\mathrm{2}/\partial V_1$ and shot noise $S_2$ can serve as two complementary electrical measurements required to  fully characterize quasiparticle transport. We apply this paradigm to explore the nonlocal response in InAs NW-based NSN devices. The outline of the experiment is depicted in Fig.~\ref{band_dia}b. A semiconducting InAs nanowire is equipped with an S terminal, made of Al, in the middle and two N terminals, made of Ti/Au bilayer, on the sides. In essence, this device represents two back-to-back N-NW-S junctions sharing the same S terminal. We study two similar devices NSN-I and NSN-II which have  the width of S terminal equal to $w=200~\mathrm{nm}$ and $w=300~\mathrm{nm}$, respectively. Note the absence of the plunger gates typically used to define the quantum dots~\cite{Hofstetter_2009,DasDas2012,Deng1557} or tunnel barriers~\cite{Albrecht2016} adjacent to the S-terminal. In addition, for all contacts in-situ Ar milling was applied before the evaporation in order to improve the semiconductor/metal interface quality.  All of this enables better coupling of the sub-gap states to the normal conducting regions. As a result, the resistances of the individual N-NW-S junctions are mainly determined by disorder scattering rather than unintentional interface reflectivity. This is confirmed by smooth gate voltage characteristics and universal diffusive value of the shot noise Fano factor in the normal state, see the data of the related experiment in Ref.~\cite{Denisov2020}. Throughout the experiments the S terminal is grounded, terminal $\mathrm{N_1}$ is current biased.  The terminal $\mathrm{N_2}$ is DC floating throughout the experiment, which allows to access the differential resistances $R_{ij}\equiv \partial V_i/\partial I_j$ in a quasi-four-terminal configuration excluding the wiring contributions. The differential conductances are obtained by inverting the measured resistance matrix, see the Supplemental Material for the details. In present experiment the non-diagonal elements are much smaller than the diagonal ones, so that approximate relations $G_{ii} \approx R_{ii}^{-1}$ and $G_{ij} \approx - R_{ij} (R_{11}R_{22})^{-1}$ hold within a few percent accuracy. Experiments are performed at bath temperatures of $T=$\,120-150\,mK unless stated otherwise.

\begin{figure}
 	\begin{center}
 		\includegraphics[width=1\columnwidth]{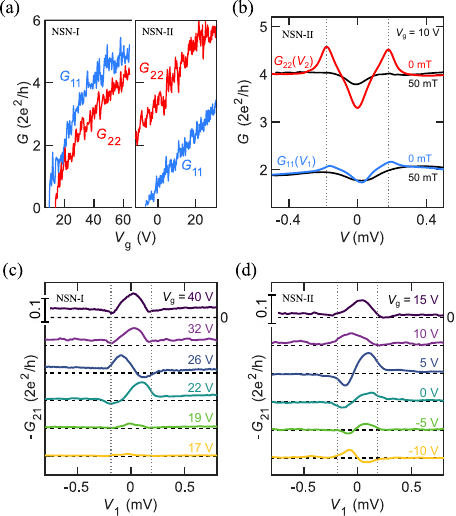}
 	\end{center}
 	\caption{(a)~Zero-bias diagonal conductances versus  $V_\mathrm{g}$ for both devices in zero magnetic field. (b)~Diagonal differential conductance as a function of bias voltage in zero and high enough to suppress superconductivity magnetic fields. Dotted lines show position of the superconducting gap. (c,~d)~ $B=0$ nonlocal differential conductance as a function of $V_\mathrm{1}$ for several $V_\mathrm{g}$ values. The curves are vertically offset for clarity with zero level shown by the dashed lines. Bars indicate the ordinate scale common for all $V_\mathrm{g}$. }
 	\label{fig2}
 \end{figure} 

Back-gate voltage ($V_\mathrm{g}$) dependencies of the linear response diagonal conductances are shown in Fig.~\ref{fig2}a. $G_{ii}$ fall in the range of a few conductance quanta and exhibit a usual sublinear increase with $V_\mathrm{g}$ accompanied by universal mesoscopic fluctuations. Standard procedure~\cite{Ford_2009} gives a field-effect mobility of $\sim300\,\mathrm{cm^2/Vs}$ underestimated because of the field screening by contacts. Impact of a superconducting proximity effect on $G_{ii}$ is similar for both devices and all $V_\mathrm{g}$ used, typical data shown in Fig.~\ref{fig2}b. In zero magnetic field $B$ a moderate zero-bias minimum is seen surrounded by maxima at  $V=\pm\Delta/e$, where $\Delta=180\,\mu$eV is the Al superconducting gap determined independently from the critical temperature, see Supplemental Material. In a uniform magnetic field of  $B=50\,$mT obtained with a superconducting solenoid, oriented perpendicularly to the substrate and high enough to suppress the superconductivity the minimum weakens and the maxima disappear, whereas the above-gap conductance remains unchanged. Overall, this is a standard for coherent diffusive NS junctions re-entrant behaviour~\cite{PhysRevB.56.14108,Courtois1999}, with a minor effect of the interface reflectivity and/or Coulomb effects~\cite{Beenakker_1997}. 

The non-diagonal conductance probes quasiparticle transport via InAs-NW section underneath the S-terminal~\cite{PhysRevB.97.045421,PhysRevLett.124.036802,PhysRevLett.124.036801,pan2020threeterminal,PhysRevLett.124.036802,puglia2020closing} and its bias dependence turns out much less universal. In Figs.~\ref{fig2}c and~\ref{fig2}d we plot $-G_\mathrm{21}$, having in mind that the negative sign corresponds to normal transmission. At sub-gap biases very different behaviour of $-G_\mathrm{21}$ can be observed depending on $V_\mathrm{g}$, from almost symmetric with zero-bias  maximum, see $V_\mathrm{g}=40\,$V data in device NSN-I, to strongly anti-symmetric with sign inversion, see $V_\mathrm{g}\leq5\,$V data in device NSN-II. By contrast, in the normal state $G_\mathrm{21}$ is negative, featureless and consistent with a current transfer length of $\sim100\,$nm determined by a residual interface reflectivity, see the Supplemental Material. Figs.~\ref{fig2}c and~\ref{fig2}d show that at sub-gap biases in the superconducting state $|G_\mathrm{21}|$ strongly increases compared to its above-gap values, which is expected since quasiparticles are forbidden to enter the superconductor. The $G_\mathrm{21}$ is much smaller than $e^2/h$ and occasionally changes sign, implying that $|\Sigma\mathcal{T}_{-}|\ll1$ and fluctuates around zero as a function of energy and chemical potential. Below we present the nonlocal shot noise experiment that uncovers charge-neutral origin of the quasiparticle transport.

\begin{figure}
 	\begin{center}
 		\includegraphics[width=1\columnwidth]{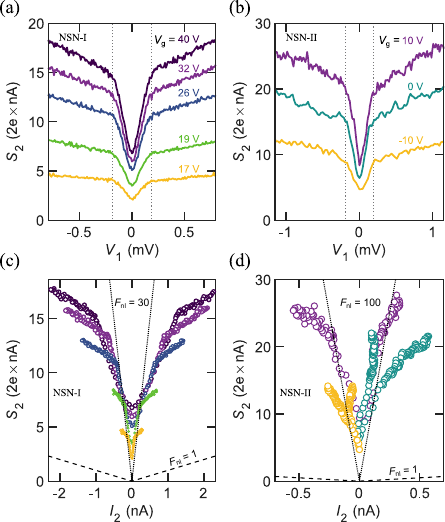}
 	\end{center}
 	\caption{(a,~b)~Measured current noise spectral density in the right lead as a function of bias voltage on the left one. Dotted lines show positions of the superconducting gap. (c,~d)~Measured current noise spectral density in the right lead as a function of the nonlocal current $I_\mathrm{2}$. Symbols have the same colour as the lines in panels (a,~b) for the respective $V_\mathrm{g}$. Guide lines with Fano-factor $F_\mathrm{nl}=1$ and $F_\mathrm{nl}\gg1$ are plotted as dashed and dotted lines correspondingly. }
 	\label{fig3}
 \end{figure} 

The layout of the shot noise measurement in a nonlocal configuration is sketched in Fig.~\ref{band_dia}b. The current fluctuations are picked up at the floating terminal $\mathrm{N_2}$ in response to the current between the biased terminal $\mathrm{N_1}$ and the grounded S-terminal. Plotted as a function of $V_1$ all the data in both devices feature the same qualitative behaviour shown in Figs.~\ref{fig3}a and~\ref{fig3}b. The shot noise spectral density $S_2$ starts from the Johnson-Nyquist equilibrium value $4k_BTG_{22}$ at zero bias~\cite{Anantram_1996} and increases almost symmetrically and linear with $V_1$ showing a pronounced downward kink at the gap edges $|V_1|=\Delta/e$ marked by vertical dashed lines. Above the gap the slope drops down consistent with the fact that transmission probability diminishes as soon as the quasiparticles can sink in the superconductor. Nonlocal noise is more informative than the usual two-terminal noise in NS structures~\cite{Jehl2000,PhysRevLett.84.3398,PhysRevB.72.024501,DasDas2012}, which exhibits only a minor reduction in the presence of sub-gap density of states~\cite{Tikhonov_PRL_2016}. According to the Eq.~(\ref{eq2}), neglecting the contributions of $\mathcal{T}_{-}^2$ the slope $dS_2/dV_1$ is determined by $\mathcal{T}_{+}$ and allows to evaluate the linear response thermal conductance $G_\mathrm{th}\equiv G_\mathrm{th}^0\Sigma\mathcal{T}_{+}$. We have $0.3<G_\mathrm{th}/G_\mathrm{th}^0<1.6$, where $G_\mathrm{th}^0 = \mathcal{L}Te^2/h$ is the thermal conductance quantum and $\mathcal{L}$ is the Lorenz number. This estimate of $G_\mathrm{th}$ is legitimate provided $|\mathcal{T}_{-}|\ll1$, i.e. if ballistic transmission is suppressed by disorder scattering~\cite{Kopnin_2004,Haim_2019}, as in our devices, or by structure geometry~\cite{Laeven_2020}. Note that our results manifest a drastic violation of the Wiedemann-Franz law~\cite{Franz_1853} since a zero resistance of the NW region covered by the superconductor coexists with a relatively small thermal conductance via the sub-gap states.

Next we analyse the same data in terms of the nonlocal Fano factor where the nonlocal current is obtained via $I_\mathrm{2}=-G_{22}V_\mathrm{2}$. Here we use that the electric current caused by non-equilibrium quasiparticles is compensated by an extra current flowing in the opposite direction, so that the net current is zero in the floating configuration. This extra current flows near the Fermi level and is noiseless, since the junction $\mathrm{N_2}$-NW-S remains essentially unbiased throughout the experiment, $|V_2|<k_BT/e$, see the Supplemental Material. Figs.~\ref{fig3}c and~\ref{fig3}d plot $S_2$  vs $I_\mathrm{2}$ for both devices. Two main features are evident. First, the symmetry inherent to $S_2$ vs $V_1$ data is in many cases lost here, since $I_\mathrm{2}$ is not an anti-symmetric function of $V_1$. Second, the noise slope corresponds to nonlocal Fano factor values in the range $30\lesssim F_\mathrm{nl}\lesssim100$, as shown by the dotted guide lines. Such giant values of $F_\mathrm{nl}$ rule out a heretical interpretation that normal quasiparticle scattering from a poor quality Al/InAs interface is the main source of nonlocal signals, that would correspond to $F_\mathrm{nl}\leq1$, see the dashed guide line.  

\begin{figure}
 	\begin{center}
 		\includegraphics[width=1\columnwidth]{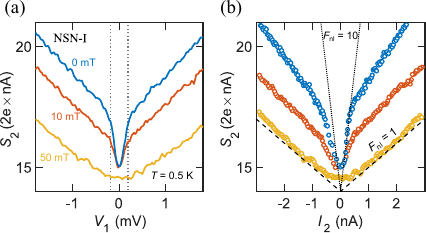}
 	\end{center}
 	\caption{(a)~Evolution of the nonlocal noise spectral density $S_{\mathrm{2}}(V_{\mathrm{1}})$ in magnetic field and $T=0.5~\mathrm{K}$. Dotted lines show positions of the superconducting gap. (b)~Evolution of the nonlocal noise spectral density $S_{\mathrm{2}}(I_{\mathrm{2}})$ in magnetic field. Symbols have the same colour as the lines in panel (a) for the respective $B$. Guide lines with Fano-factor $F_\mathrm{nl}=1$ and $F_\mathrm{nl}\gg1$ are plotted as dashed and dotted lines correspondingly. }
 	\label{fig4}
 \end{figure} 

It is convenient to define the average charge of transmitted quasiparticles as the ratio between the transmitted charge and total number of transmitted quasiparticles $\langle q_\mathrm{T}\rangle = \Sigma\mathcal{T}_{-}/\Sigma\mathcal{T}_{+}$. The value of $\langle q_\mathrm{T}\rangle=1$ corresponds to the case when quasiparticles conserve their charge during the transmission process, whereas in the case  $\langle q_\mathrm{T}\rangle=-1$ they invert the charge. As follows from the Eqs.~(\ref{eq1}-\ref{eq2}),  $|\langle q_\mathrm{T}\rangle| < 1/F_\mathrm{nl}$, i.e. the observation of a giant Fano factor implies nearly charge-neutral nonlocal quasiparticle transport with $|\langle q_\mathrm{T}\rangle|\ll1$. This can be easily understood in case of a metallic diffusive NW covered by a superconductor with a transparent interface. Traversing the proximity region a quasiparticle experiences a number of ARs given on the average by $\langle N_\mathrm{AR} \rangle= (w/d)^2$, where $w$ is the width of the S-terminal and $d\approx100\,$nm is the diameter of the NW. Given the randomness of diffusion the mean-square fluctuation is $\sqrt{\langle \delta N_\mathrm{AR}^2\rangle}=\sqrt{\langle N_\mathrm{AR}\rangle}\geq2$, so that the parity of $N_\mathrm{AR}$ and thus the sign of the transmitted charge are completely uncertain. More rigorously, for the device NSN-II we find $|\langle q_\mathrm{T}\rangle|<0.01$, meaning that it takes at least a hundred quasiparticles to transmit a unit of elementary charge.  

Our observations in a trivial superconducting phase have common features with the predicted nonlocal response at the topological phase transition in Majorana NWs. Here, even in presence of a moderate disorder, a finite transmission occurs in just one eigenchannel with $T_{21}=A_{21}=1/4$ and results in a pure heat transport characterized by a universal peak of the thermal conductance $G_\mathrm{th}=G_\mathrm{th}^0/2$~\cite{PhysRevLett.106.057001}. While comparable in absolute value, $G_\mathrm{th}$ in the present experiment demonstrates a monotonic dependence on $V_\mathrm{g}$. The nonlocal charge response at the topological transition restores at a finite bias, $G_{21}\propto V_1$, owing to the energy-dependence of the transmission probabilities, known as the Andreev rectification~\cite{PhysRevB.97.045421}. Similar transport features are occasionally observable in Figs.~\ref{fig2}c and~\ref{fig2}d, originating from mesoscopic fluctuations of $G_{21}$ around zero. This suggests that, unlike the peak in $G_\mathrm{th}$,  Andreev rectification is not a unique signature of the topological transition, see also Ref.~\cite{puglia2020closing}. 

As a final step, we demonstrate a crossover from nearly charge-neutral to normal nonlocal quasiparticle transport in a magnetic field in the device NSN-I.  Fig.~\ref{fig4}a shows the evolution of $S_2$ vs $V_1$, taken  at a bath temperature of 0.5\,K. The shot noise gradually diminishes at increasing $B$-field and the kink at the gap edge disappears concurrently with a transition of the Al to the normal state, see the $B=50\,$mT trace. Plotted as a function of $I_\mathrm{2}$ in Fig.~\ref{fig4}b this data reveals a transition from the giant noise at sub-gap energies in the $B=0$ superconducting state to the Poissonian noise in the normal state, see the dashed guide line. In $B=0$ we find $F_\mathrm{nl}\sim10$, considerably diminished compared to Fig.~\ref{fig3}c as a result of thermal smearing. The Poissonian noise $F_\mathrm{nl}\approx1$ in the normal state corresponds to $A_{21}=0$ and  $T_{21}\ll1$, as a result of residual interface scattering. Notably, in the superconducting state the slope corresponds to $F_\mathrm{nl}>1$ even beyond the kink, see also Figs.~\ref{fig3}c and~\ref{fig3}d, since the probability of the AR remains finite above the gap~\cite{tinkham2004introduction}.

In summary, we performed nonlocal transport and noise experiments in InAs NW-based hybrid NSN devices. Such a combination of all-electrical measurements allows to estimate the thermal conductance and reveals a predominantly charge-neutral origin of the nonlocal response. We expect that our approach will be generally useful for the studies of non-equilibrium proximity superconductivity, including unequivocal identification of the topological phase transition in Majorana devices.

We are grateful to A.P. Higginbotham, T.M. Klapwijk, A.S. Mel'nikov  and K.E. Nagaev for helpful discussions. This work was financially supported by the RSF project 19-12-00326 (fabrication, experiments in NSN-I device) and RFBR project 19-02-00898 (experiments in NSN-II device). Theoretical framework was developed under the state task of the ISSP RAS. Work at TUM was supported by the Deutsche Forschungsgemeinschaft (DFG) via project KO-4005/5-1 and Germany's Excellence Strategy-EXC-2111-390814868 (Munich Center for Quantum Science and Technology, MCQST).

\renewcommand*{\bibfont}{\small}

\begin{thebibliography}{47}%
\makeatletter
\providecommand \@ifxundefined [1]{%
 \@ifx{#1\undefined}
}%
\providecommand \@ifnum [1]{%
 \ifnum #1\expandafter \@firstoftwo
 \else \expandafter \@secondoftwo
 \fi
}%
\providecommand \@ifx [1]{%
 \ifx #1\expandafter \@firstoftwo
 \else \expandafter \@secondoftwo
 \fi
}%
\providecommand \natexlab [1]{#1}%
\providecommand \enquote  [1]{``#1''}%
\providecommand \bibnamefont  [1]{#1}%
\providecommand \bibfnamefont [1]{#1}%
\providecommand \citenamefont [1]{#1}%
\providecommand \href@noop [0]{\@secondoftwo}%
\providecommand \href [0]{\begingroup \@sanitize@url \@href}%
\providecommand \@href[1]{\@@startlink{#1}\@@href}%
\providecommand \@@href[1]{\endgroup#1\@@endlink}%
\providecommand \@sanitize@url [0]{\catcode `\\12\catcode `\$12\catcode
  `\&12\catcode `\#12\catcode `\^12\catcode `\_12\catcode `\%12\relax}%
\providecommand \@@startlink[1]{}%
\providecommand \@@endlink[0]{}%
\providecommand \url  [0]{\begingroup\@sanitize@url \@url }%
\providecommand \@url [1]{\endgroup\@href {#1}{\urlprefix }}%
\providecommand \urlprefix  [0]{URL }%
\providecommand \Eprint [0]{\href }%
\providecommand \doibase [0]{https://doi.org/}%
\providecommand \selectlanguage [0]{\@gobble}%
\providecommand \bibinfo  [0]{\@secondoftwo}%
\providecommand \bibfield  [0]{\@secondoftwo}%
\providecommand \translation [1]{[#1]}%
\providecommand \BibitemOpen [0]{}%
\providecommand \bibitemStop [0]{}%
\providecommand \bibitemNoStop [0]{.\EOS\space}%
\providecommand \EOS [0]{\spacefactor3000\relax}%
\providecommand \BibitemShut  [1]{\csname bibitem#1\endcsname}%
\let\auto@bib@innerbib\@empty
\bibitem [{\citenamefont {den Hartog}\ \emph {et~al.}(1996)\citenamefont {den
  Hartog}, \citenamefont {Kapteyn}, \citenamefont {van Wees}, \citenamefont
  {Klapwijk},\ and\ \citenamefont {Borghs}}]{PhysRevLett.77.4954}%
  \BibitemOpen
  \bibfield  {author} {\bibinfo {author} {\bibfnamefont {S.~G.}\ \bibnamefont
  {den Hartog}}, \bibinfo {author} {\bibfnamefont {C.~M.~A.}\ \bibnamefont
  {Kapteyn}}, \bibinfo {author} {\bibfnamefont {B.~J.}\ \bibnamefont {van
  Wees}}, \bibinfo {author} {\bibfnamefont {T.~M.}\ \bibnamefont {Klapwijk}},\
  and\ \bibinfo {author} {\bibfnamefont {G.}~\bibnamefont {Borghs}},\
  }\bibfield  {title} {\bibinfo {title} {Transport in multiterminal
  normal-superconductor devices: Reciprocity relations, negative and nonlocal
  resistances, and reentrance of the proximity effect},\ }\href
  {https://doi.org/10.1103/PhysRevLett.77.4954} {\bibfield  {journal} {\bibinfo
   {journal} {Phys. Rev. Lett.}\ }\textbf {\bibinfo {volume} {77}},\ \bibinfo
  {pages} {4954} (\bibinfo {year} {1996})}\BibitemShut {NoStop}%
\bibitem [{\citenamefont {Lee}\ \emph {et~al.}(2017)\citenamefont {Lee},
  \citenamefont {Huang}, \citenamefont {Efetov}, \citenamefont {Wei},
  \citenamefont {Hart}, \citenamefont {Taniguchi}, \citenamefont {Watanabe},
  \citenamefont {Yacoby},\ and\ \citenamefont {Kim}}]{Lee_2017}%
  \BibitemOpen
  \bibfield  {author} {\bibinfo {author} {\bibfnamefont {G.-H.}\ \bibnamefont
  {Lee}}, \bibinfo {author} {\bibfnamefont {K.-F.}\ \bibnamefont {Huang}},
  \bibinfo {author} {\bibfnamefont {D.~K.}\ \bibnamefont {Efetov}}, \bibinfo
  {author} {\bibfnamefont {D.~S.}\ \bibnamefont {Wei}}, \bibinfo {author}
  {\bibfnamefont {S.}~\bibnamefont {Hart}}, \bibinfo {author} {\bibfnamefont
  {T.}~\bibnamefont {Taniguchi}}, \bibinfo {author} {\bibfnamefont
  {K.}~\bibnamefont {Watanabe}}, \bibinfo {author} {\bibfnamefont
  {A.}~\bibnamefont {Yacoby}},\ and\ \bibinfo {author} {\bibfnamefont
  {P.}~\bibnamefont {Kim}},\ }\bibfield  {title} {\bibinfo {title} {Inducing
  superconducting correlation in quantum hall edge states},\ }\href
  {https://doi.org/10.1038/nphys4084} {\bibfield  {journal} {\bibinfo
  {journal} {Nature Physics}\ }\textbf {\bibinfo {volume} {13}},\ \bibinfo
  {pages} {693} (\bibinfo {year} {2017})}\BibitemShut {NoStop}%
\bibitem [{\citenamefont {Deng}\ \emph {et~al.}(2018)\citenamefont {Deng},
  \citenamefont {Vaitiek\ifmmode~\dot{e}\else \.{e}\fi{}nas}, \citenamefont
  {Prada}, \citenamefont {San-Jose}, \citenamefont {Nyg\aa{}rd}, \citenamefont
  {Krogstrup}, \citenamefont {Aguado},\ and\ \citenamefont
  {Marcus}}]{PhysRevB.98.085125}%
  \BibitemOpen
  \bibfield  {author} {\bibinfo {author} {\bibfnamefont {M.-T.}\ \bibnamefont
  {Deng}}, \bibinfo {author} {\bibfnamefont {S.}~\bibnamefont
  {Vaitiek\ifmmode~\dot{e}\else \.{e}\fi{}nas}}, \bibinfo {author}
  {\bibfnamefont {E.}~\bibnamefont {Prada}}, \bibinfo {author} {\bibfnamefont
  {P.}~\bibnamefont {San-Jose}}, \bibinfo {author} {\bibfnamefont
  {J.}~\bibnamefont {Nyg\aa{}rd}}, \bibinfo {author} {\bibfnamefont
  {P.}~\bibnamefont {Krogstrup}}, \bibinfo {author} {\bibfnamefont
  {R.}~\bibnamefont {Aguado}},\ and\ \bibinfo {author} {\bibfnamefont {C.~M.}\
  \bibnamefont {Marcus}},\ }\bibfield  {title} {\bibinfo {title} {Nonlocality
  of majorana modes in hybrid nanowires},\ }\href
  {https://doi.org/10.1103/PhysRevB.98.085125} {\bibfield  {journal} {\bibinfo
  {journal} {Phys. Rev. B}\ }\textbf {\bibinfo {volume} {98}},\ \bibinfo
  {pages} {085125} (\bibinfo {year} {2018})}\BibitemShut {NoStop}%
\bibitem [{\citenamefont {Rosdahl}\ \emph {et~al.}(2018)\citenamefont
  {Rosdahl}, \citenamefont {Vuik}, \citenamefont {Kjaergaard},\ and\
  \citenamefont {Akhmerov}}]{PhysRevB.97.045421}%
  \BibitemOpen
  \bibfield  {author} {\bibinfo {author} {\bibfnamefont {T.~O.}\ \bibnamefont
  {Rosdahl}}, \bibinfo {author} {\bibfnamefont {A.}~\bibnamefont {Vuik}},
  \bibinfo {author} {\bibfnamefont {M.}~\bibnamefont {Kjaergaard}},\ and\
  \bibinfo {author} {\bibfnamefont {A.~R.}\ \bibnamefont {Akhmerov}},\
  }\bibfield  {title} {\bibinfo {title} {Andreev rectifier: A nonlocal
  conductance signature of topological phase transitions},\ }\href
  {https://doi.org/10.1103/PhysRevB.97.045421} {\bibfield  {journal} {\bibinfo
  {journal} {Phys. Rev. B}\ }\textbf {\bibinfo {volume} {97}},\ \bibinfo
  {pages} {045421} (\bibinfo {year} {2018})}\BibitemShut {NoStop}%
\bibitem [{\citenamefont {Lai}\ \emph {et~al.}(2019)\citenamefont {Lai},
  \citenamefont {Sau},\ and\ \citenamefont {Das~Sarma}}]{PhysRevB.100.045302}%
  \BibitemOpen
  \bibfield  {author} {\bibinfo {author} {\bibfnamefont {Y.-H.}\ \bibnamefont
  {Lai}}, \bibinfo {author} {\bibfnamefont {J.~D.}\ \bibnamefont {Sau}},\ and\
  \bibinfo {author} {\bibfnamefont {S.}~\bibnamefont {Das~Sarma}},\ }\bibfield
  {title} {\bibinfo {title} {Presence versus absence of end-to-end nonlocal
  conductance correlations in majorana nanowires: Majorana bound states versus
  andreev bound states},\ }\href {https://doi.org/10.1103/PhysRevB.100.045302}
  {\bibfield  {journal} {\bibinfo  {journal} {Phys. Rev. B}\ }\textbf {\bibinfo
  {volume} {100}},\ \bibinfo {pages} {045302} (\bibinfo {year}
  {2019})}\BibitemShut {NoStop}%
\bibitem [{\citenamefont {Danon}\ \emph {et~al.}(2020)\citenamefont {Danon},
  \citenamefont {Hellenes}, \citenamefont {Hansen}, \citenamefont {Casparis},
  \citenamefont {Higginbotham},\ and\ \citenamefont
  {Flensberg}}]{PhysRevLett.124.036801}%
  \BibitemOpen
  \bibfield  {author} {\bibinfo {author} {\bibfnamefont {J.}~\bibnamefont
  {Danon}}, \bibinfo {author} {\bibfnamefont {A.~B.}\ \bibnamefont {Hellenes}},
  \bibinfo {author} {\bibfnamefont {E.~B.}\ \bibnamefont {Hansen}}, \bibinfo
  {author} {\bibfnamefont {L.}~\bibnamefont {Casparis}}, \bibinfo {author}
  {\bibfnamefont {A.~P.}\ \bibnamefont {Higginbotham}},\ and\ \bibinfo {author}
  {\bibfnamefont {K.}~\bibnamefont {Flensberg}},\ }\bibfield  {title} {\bibinfo
  {title} {Nonlocal conductance spectroscopy of andreev bound states: Symmetry
  relations and bcs charges},\ }\href
  {https://doi.org/10.1103/PhysRevLett.124.036801} {\bibfield  {journal}
  {\bibinfo  {journal} {Phys. Rev. Lett.}\ }\textbf {\bibinfo {volume} {124}},\
  \bibinfo {pages} {036801} (\bibinfo {year} {2020})}\BibitemShut {NoStop}%
\bibitem [{\citenamefont {Pan}\ \emph {et~al.}(2021)\citenamefont {Pan},
  \citenamefont {Sau},\ and\ \citenamefont {Das~Sarma}}]{pan2020threeterminal}%
  \BibitemOpen
  \bibfield  {author} {\bibinfo {author} {\bibfnamefont {H.}~\bibnamefont
  {Pan}}, \bibinfo {author} {\bibfnamefont {J.~D.}\ \bibnamefont {Sau}},\ and\
  \bibinfo {author} {\bibfnamefont {S.}~\bibnamefont {Das~Sarma}},\ }\bibfield
  {title} {\bibinfo {title} {Three-terminal nonlocal conductance in majorana
  nanowires: Distinguishing topological and trivial in realistic systems with
  disorder and inhomogeneous potential},\ }\href
  {https://doi.org/10.1103/PhysRevB.103.014513} {\bibfield  {journal} {\bibinfo
   {journal} {Phys. Rev. B}\ }\textbf {\bibinfo {volume} {103}},\ \bibinfo
  {pages} {014513} (\bibinfo {year} {2021})}\BibitemShut {NoStop}%
\bibitem [{\citenamefont {Zhao}\ \emph {et~al.}(2020)\citenamefont {Zhao},
  \citenamefont {Arnault}, \citenamefont {Bondarev}, \citenamefont
  {Seredinski}, \citenamefont {Larson}, \citenamefont {Draelos}, \citenamefont
  {Li}, \citenamefont {Watanabe}, \citenamefont {Taniguchi}, \citenamefont
  {Amet}, \citenamefont {Baranger},\ and\ \citenamefont
  {Finkelstein}}]{Zhao_2020}%
  \BibitemOpen
  \bibfield  {author} {\bibinfo {author} {\bibfnamefont {L.}~\bibnamefont
  {Zhao}}, \bibinfo {author} {\bibfnamefont {E.~G.}\ \bibnamefont {Arnault}},
  \bibinfo {author} {\bibfnamefont {A.}~\bibnamefont {Bondarev}}, \bibinfo
  {author} {\bibfnamefont {A.}~\bibnamefont {Seredinski}}, \bibinfo {author}
  {\bibfnamefont {T.~F.~Q.}\ \bibnamefont {Larson}}, \bibinfo {author}
  {\bibfnamefont {A.~W.}\ \bibnamefont {Draelos}}, \bibinfo {author}
  {\bibfnamefont {H.}~\bibnamefont {Li}}, \bibinfo {author} {\bibfnamefont
  {K.}~\bibnamefont {Watanabe}}, \bibinfo {author} {\bibfnamefont
  {T.}~\bibnamefont {Taniguchi}}, \bibinfo {author} {\bibfnamefont
  {F.}~\bibnamefont {Amet}}, \bibinfo {author} {\bibfnamefont {H.~U.}\
  \bibnamefont {Baranger}},\ and\ \bibinfo {author} {\bibfnamefont
  {G.}~\bibnamefont {Finkelstein}},\ }\bibfield  {title} {\bibinfo {title}
  {Interference of chiral andreev edge states},\ }\bibfield  {journal}
  {\bibinfo  {journal} {Nature Physics}\ }\href
  {https://doi.org/10.1038/s41567-020-0898-5} {10.1038/s41567-020-0898-5}
  (\bibinfo {year} {2020})\BibitemShut {NoStop}%
\bibitem [{\citenamefont {Lutchyn}\ \emph {et~al.}(2010)\citenamefont
  {Lutchyn}, \citenamefont {Sau},\ and\ \citenamefont
  {Das~Sarma}}]{PhysRevLett.105.077001}%
  \BibitemOpen
  \bibfield  {author} {\bibinfo {author} {\bibfnamefont {R.~M.}\ \bibnamefont
  {Lutchyn}}, \bibinfo {author} {\bibfnamefont {J.~D.}\ \bibnamefont {Sau}},\
  and\ \bibinfo {author} {\bibfnamefont {S.}~\bibnamefont {Das~Sarma}},\
  }\bibfield  {title} {\bibinfo {title} {Majorana fermions and a topological
  phase transition in semiconductor-superconductor heterostructures},\ }\href
  {https://doi.org/10.1103/PhysRevLett.105.077001} {\bibfield  {journal}
  {\bibinfo  {journal} {Phys. Rev. Lett.}\ }\textbf {\bibinfo {volume} {105}},\
  \bibinfo {pages} {077001} (\bibinfo {year} {2010})}\BibitemShut {NoStop}%
\bibitem [{\citenamefont {Oreg}\ \emph {et~al.}(2010)\citenamefont {Oreg},
  \citenamefont {Refael},\ and\ \citenamefont {von
  Oppen}}]{PhysRevLett.105.177002}%
  \BibitemOpen
  \bibfield  {author} {\bibinfo {author} {\bibfnamefont {Y.}~\bibnamefont
  {Oreg}}, \bibinfo {author} {\bibfnamefont {G.}~\bibnamefont {Refael}},\ and\
  \bibinfo {author} {\bibfnamefont {F.}~\bibnamefont {von Oppen}},\ }\bibfield
  {title} {\bibinfo {title} {Helical liquids and majorana bound states in
  quantum wires},\ }\href {https://doi.org/10.1103/PhysRevLett.105.177002}
  {\bibfield  {journal} {\bibinfo  {journal} {Phys. Rev. Lett.}\ }\textbf
  {\bibinfo {volume} {105}},\ \bibinfo {pages} {177002} (\bibinfo {year}
  {2010})}\BibitemShut {NoStop}%
\bibitem [{\citenamefont {Akhmerov}\ \emph {et~al.}(2011)\citenamefont
  {Akhmerov}, \citenamefont {Dahlhaus}, \citenamefont {Hassler}, \citenamefont
  {Wimmer},\ and\ \citenamefont {Beenakker}}]{PhysRevLett.106.057001}%
  \BibitemOpen
  \bibfield  {author} {\bibinfo {author} {\bibfnamefont {A.~R.}\ \bibnamefont
  {Akhmerov}}, \bibinfo {author} {\bibfnamefont {J.~P.}\ \bibnamefont
  {Dahlhaus}}, \bibinfo {author} {\bibfnamefont {F.}~\bibnamefont {Hassler}},
  \bibinfo {author} {\bibfnamefont {M.}~\bibnamefont {Wimmer}},\ and\ \bibinfo
  {author} {\bibfnamefont {C.~W.~J.}\ \bibnamefont {Beenakker}},\ }\bibfield
  {title} {\bibinfo {title} {Quantized conductance at the majorana phase
  transition in a disordered superconducting wire},\ }\href
  {https://doi.org/10.1103/PhysRevLett.106.057001} {\bibfield  {journal}
  {\bibinfo  {journal} {Phys. Rev. Lett.}\ }\textbf {\bibinfo {volume} {106}},\
  \bibinfo {pages} {057001} (\bibinfo {year} {2011})}\BibitemShut {NoStop}%
\bibitem [{\citenamefont {Mourik}\ \emph {et~al.}(2012)\citenamefont {Mourik},
  \citenamefont {Zuo}, \citenamefont {Frolov}, \citenamefont {Plissard},
  \citenamefont {Bakkers},\ and\ \citenamefont {Kouwenhoven}}]{Mourik1003}%
  \BibitemOpen
  \bibfield  {author} {\bibinfo {author} {\bibfnamefont {V.}~\bibnamefont
  {Mourik}}, \bibinfo {author} {\bibfnamefont {K.}~\bibnamefont {Zuo}},
  \bibinfo {author} {\bibfnamefont {S.~M.}\ \bibnamefont {Frolov}}, \bibinfo
  {author} {\bibfnamefont {S.~R.}\ \bibnamefont {Plissard}}, \bibinfo {author}
  {\bibfnamefont {E.~P. A.~M.}\ \bibnamefont {Bakkers}},\ and\ \bibinfo
  {author} {\bibfnamefont {L.~P.}\ \bibnamefont {Kouwenhoven}},\ }\bibfield
  {title} {\bibinfo {title} {Signatures of majorana fermions in hybrid
  superconductor-semiconductor nanowire devices},\ }\href
  {https://doi.org/10.1126/science.1222360} {\bibfield  {journal} {\bibinfo
  {journal} {Science}\ }\textbf {\bibinfo {volume} {336}},\ \bibinfo {pages}
  {1003} (\bibinfo {year} {2012})}\BibitemShut {NoStop}%
\bibitem [{\citenamefont {Das}\ \emph {et~al.}(2012{\natexlab{a}})\citenamefont
  {Das}, \citenamefont {Ronen}, \citenamefont {Most}, \citenamefont {Oreg},
  \citenamefont {Heiblum},\ and\ \citenamefont {Shtrikman}}]{Das2012}%
  \BibitemOpen
  \bibfield  {author} {\bibinfo {author} {\bibfnamefont {A.}~\bibnamefont
  {Das}}, \bibinfo {author} {\bibfnamefont {Y.}~\bibnamefont {Ronen}}, \bibinfo
  {author} {\bibfnamefont {Y.}~\bibnamefont {Most}}, \bibinfo {author}
  {\bibfnamefont {Y.}~\bibnamefont {Oreg}}, \bibinfo {author} {\bibfnamefont
  {M.}~\bibnamefont {Heiblum}},\ and\ \bibinfo {author} {\bibfnamefont
  {H.}~\bibnamefont {Shtrikman}},\ }\bibfield  {title} {\bibinfo {title}
  {Zero-bias peaks and splitting in an al-inas nanowire topological
  superconductor as a signature of majorana fermions},\ }\href
  {https://doi.org/10.1038/nphys2479} {\bibfield  {journal} {\bibinfo
  {journal} {Nature Physics}\ }\textbf {\bibinfo {volume} {8}},\ \bibinfo
  {pages} {887 EP } (\bibinfo {year} {2012}{\natexlab{a}})},\ \bibinfo {note}
  {article}\BibitemShut {NoStop}%
\bibitem [{\citenamefont {Yu}\ \emph {et~al.}(2021)\citenamefont {Yu},
  \citenamefont {Chen}, \citenamefont {Gomanko}, \citenamefont {Badawy},
  \citenamefont {Bakkers}, \citenamefont {Zuo}, \citenamefont {Mourik},\ and\
  \citenamefont {Frolov}}]{yu2020nonmajorana}%
  \BibitemOpen
  \bibfield  {author} {\bibinfo {author} {\bibfnamefont {P.}~\bibnamefont
  {Yu}}, \bibinfo {author} {\bibfnamefont {J.}~\bibnamefont {Chen}}, \bibinfo
  {author} {\bibfnamefont {M.}~\bibnamefont {Gomanko}}, \bibinfo {author}
  {\bibfnamefont {G.}~\bibnamefont {Badawy}}, \bibinfo {author} {\bibfnamefont
  {E.~P. A.~M.}\ \bibnamefont {Bakkers}}, \bibinfo {author} {\bibfnamefont
  {K.}~\bibnamefont {Zuo}}, \bibinfo {author} {\bibfnamefont {V.}~\bibnamefont
  {Mourik}},\ and\ \bibinfo {author} {\bibfnamefont {S.~M.}\ \bibnamefont
  {Frolov}},\ }\bibfield  {title} {\bibinfo {title} {Non-majorana states yield
  nearly quantized conductance in proximatized nanowires},\ }\href
  {https://doi.org/10.1038/s41567-020-01107-w} {\bibfield  {journal} {\bibinfo
  {journal} {Nature Physics}\ }\textbf {\bibinfo {volume} {17}},\ \bibinfo
  {pages} {482} (\bibinfo {year} {2021})}\BibitemShut {NoStop}%
\bibitem [{\citenamefont {M\'enard}\ \emph {et~al.}(2020)\citenamefont
  {M\'enard}, \citenamefont {Anselmetti}, \citenamefont {Martinez},
  \citenamefont {Puglia}, \citenamefont {Malinowski}, \citenamefont {Lee},
  \citenamefont {Choi}, \citenamefont {Pendharkar}, \citenamefont
  {Palmstr\o{}m}, \citenamefont {Flensberg}, \citenamefont {Marcus},
  \citenamefont {Casparis},\ and\ \citenamefont
  {Higginbotham}}]{PhysRevLett.124.036802}%
  \BibitemOpen
  \bibfield  {author} {\bibinfo {author} {\bibfnamefont {G.~C.}\ \bibnamefont
  {M\'enard}}, \bibinfo {author} {\bibfnamefont {G.~L.~R.}\ \bibnamefont
  {Anselmetti}}, \bibinfo {author} {\bibfnamefont {E.~A.}\ \bibnamefont
  {Martinez}}, \bibinfo {author} {\bibfnamefont {D.}~\bibnamefont {Puglia}},
  \bibinfo {author} {\bibfnamefont {F.~K.}\ \bibnamefont {Malinowski}},
  \bibinfo {author} {\bibfnamefont {J.~S.}\ \bibnamefont {Lee}}, \bibinfo
  {author} {\bibfnamefont {S.}~\bibnamefont {Choi}}, \bibinfo {author}
  {\bibfnamefont {M.}~\bibnamefont {Pendharkar}}, \bibinfo {author}
  {\bibfnamefont {C.~J.}\ \bibnamefont {Palmstr\o{}m}}, \bibinfo {author}
  {\bibfnamefont {K.}~\bibnamefont {Flensberg}}, \bibinfo {author}
  {\bibfnamefont {C.~M.}\ \bibnamefont {Marcus}}, \bibinfo {author}
  {\bibfnamefont {L.}~\bibnamefont {Casparis}},\ and\ \bibinfo {author}
  {\bibfnamefont {A.~P.}\ \bibnamefont {Higginbotham}},\ }\bibfield  {title}
  {\bibinfo {title} {Conductance-matrix symmetries of a three-terminal hybrid
  device},\ }\href {https://doi.org/10.1103/PhysRevLett.124.036802} {\bibfield
  {journal} {\bibinfo  {journal} {Phys. Rev. Lett.}\ }\textbf {\bibinfo
  {volume} {124}},\ \bibinfo {pages} {036802} (\bibinfo {year}
  {2020})}\BibitemShut {NoStop}%
\bibitem [{\citenamefont {Puglia}\ \emph {et~al.}(2021)\citenamefont {Puglia},
  \citenamefont {Martinez}, \citenamefont {M\'enard}, \citenamefont {P\"oschl},
  \citenamefont {Gronin}, \citenamefont {Gardner}, \citenamefont {Kallaher},
  \citenamefont {Manfra}, \citenamefont {Marcus}, \citenamefont
  {Higginbotham},\ and\ \citenamefont {Casparis}}]{puglia2020closing}%
  \BibitemOpen
  \bibfield  {author} {\bibinfo {author} {\bibfnamefont {D.}~\bibnamefont
  {Puglia}}, \bibinfo {author} {\bibfnamefont {E.~A.}\ \bibnamefont
  {Martinez}}, \bibinfo {author} {\bibfnamefont {G.~C.}\ \bibnamefont
  {M\'enard}}, \bibinfo {author} {\bibfnamefont {A.}~\bibnamefont {P\"oschl}},
  \bibinfo {author} {\bibfnamefont {S.}~\bibnamefont {Gronin}}, \bibinfo
  {author} {\bibfnamefont {G.~C.}\ \bibnamefont {Gardner}}, \bibinfo {author}
  {\bibfnamefont {R.}~\bibnamefont {Kallaher}}, \bibinfo {author}
  {\bibfnamefont {M.~J.}\ \bibnamefont {Manfra}}, \bibinfo {author}
  {\bibfnamefont {C.~M.}\ \bibnamefont {Marcus}}, \bibinfo {author}
  {\bibfnamefont {A.~P.}\ \bibnamefont {Higginbotham}},\ and\ \bibinfo {author}
  {\bibfnamefont {L.}~\bibnamefont {Casparis}},\ }\bibfield  {title} {\bibinfo
  {title} {Closing of the induced gap in a hybrid superconductor-semiconductor
  nanowire},\ }\href {https://doi.org/10.1103/PhysRevB.103.235201} {\bibfield
  {journal} {\bibinfo  {journal} {Phys. Rev. B}\ }\textbf {\bibinfo {volume}
  {103}},\ \bibinfo {pages} {235201} (\bibinfo {year} {2021})}\BibitemShut
  {NoStop}%
\bibitem [{\citenamefont {Andreev}(1964)}]{Andreev}%
  \BibitemOpen
  \bibfield  {author} {\bibinfo {author} {\bibfnamefont {A.~F.}\ \bibnamefont
  {Andreev}},\ }\bibfield  {title} {\bibinfo {title} {Thermal conductivity of
  the intermediate state of superconductors},\ }\href@noop {} {\bibfield
  {journal} {\bibinfo  {journal} {JETP Letters}\ }\textbf {\bibinfo {volume}
  {46}},\ \bibinfo {pages} {1823} (\bibinfo {year} {1964})}\BibitemShut
  {NoStop}%
\bibitem [{\citenamefont {Kopnin}\ \emph {et~al.}(2004)\citenamefont {Kopnin},
  \citenamefont {Mel'nikov},\ and\ \citenamefont {Vinokur}}]{Kopnin_2004}%
  \BibitemOpen
  \bibfield  {author} {\bibinfo {author} {\bibfnamefont {N.~B.}\ \bibnamefont
  {Kopnin}}, \bibinfo {author} {\bibfnamefont {A.~S.}\ \bibnamefont
  {Mel'nikov}},\ and\ \bibinfo {author} {\bibfnamefont {V.~M.}\ \bibnamefont
  {Vinokur}},\ }\bibfield  {title} {\bibinfo {title} {Reentrant localization of
  single-particle transport in disordered andreev wires},\ }\href
  {https://doi.org/10.1103/PhysRevB.70.075310} {\bibfield  {journal} {\bibinfo
  {journal} {Phys. Rev. B}\ }\textbf {\bibinfo {volume} {70}},\ \bibinfo
  {pages} {075310} (\bibinfo {year} {2004})}\BibitemShut {NoStop}%
\bibitem [{\citenamefont {Tinkham}(2004)}]{tinkham2004introduction}%
  \BibitemOpen
  \bibfield  {author} {\bibinfo {author} {\bibfnamefont {M.}~\bibnamefont
  {Tinkham}},\ }\href {https://books.google.ru/books?id=VpUk3NfwDIkC} {\emph
  {\bibinfo {title} {Introduction to Superconductivity}}},\ Dover Books on
  Physics Series\ (\bibinfo  {publisher} {Dover Publications, NewYork},\
  \bibinfo {year} {2004})\BibitemShut {NoStop}%
\bibitem [{\citenamefont {Heikkilä}\ \emph {et~al.}(2019)\citenamefont
  {Heikkilä}, \citenamefont {Silaev}, \citenamefont {Virtanen},\ and\
  \citenamefont {Bergeret}}]{Heikkila_2019}%
  \BibitemOpen
  \bibfield  {author} {\bibinfo {author} {\bibfnamefont {T.~T.}\ \bibnamefont
  {Heikkilä}}, \bibinfo {author} {\bibfnamefont {M.}~\bibnamefont {Silaev}},
  \bibinfo {author} {\bibfnamefont {P.}~\bibnamefont {Virtanen}},\ and\
  \bibinfo {author} {\bibfnamefont {F.~S.}\ \bibnamefont {Bergeret}},\
  }\bibfield  {title} {\bibinfo {title} {Thermal, electric and spin transport
  in superconductor/ferromagnetic-insulator structures},\ }\href
  {https://doi.org/10.1016/j.progsurf.2019.100540} {\bibfield  {journal}
  {\bibinfo  {journal} {Progress in Surface Science}\ }\textbf {\bibinfo
  {volume} {94}},\ \bibinfo {pages} {100540} (\bibinfo {year}
  {2019})}\BibitemShut {NoStop}%
\bibitem [{\citenamefont {Tikhonov}\ \emph
  {et~al.}(2016{\natexlab{a}})\citenamefont {Tikhonov}, \citenamefont
  {Shovkun}, \citenamefont {Ercolani}, \citenamefont {Rossella}, \citenamefont
  {Rocci}, \citenamefont {Sorba}, \citenamefont {Roddaro},\ and\ \citenamefont
  {Khrapai}}]{Tikhonov2016}%
  \BibitemOpen
  \bibfield  {author} {\bibinfo {author} {\bibfnamefont {E.~S.}\ \bibnamefont
  {Tikhonov}}, \bibinfo {author} {\bibfnamefont {D.~V.}\ \bibnamefont
  {Shovkun}}, \bibinfo {author} {\bibfnamefont {D.}~\bibnamefont {Ercolani}},
  \bibinfo {author} {\bibfnamefont {F.}~\bibnamefont {Rossella}}, \bibinfo
  {author} {\bibfnamefont {M.}~\bibnamefont {Rocci}}, \bibinfo {author}
  {\bibfnamefont {L.}~\bibnamefont {Sorba}}, \bibinfo {author} {\bibfnamefont
  {S.}~\bibnamefont {Roddaro}},\ and\ \bibinfo {author} {\bibfnamefont {V.~S.}\
  \bibnamefont {Khrapai}},\ }\bibfield  {title} {\bibinfo {title} {Local noise
  in a diffusive conductor},\ }\href {https://doi.org/10.1038/srep30621}
  {\bibfield  {journal} {\bibinfo  {journal} {Scientific Reports}\ }\textbf
  {\bibinfo {volume} {6}},\ \bibinfo {pages} {30621 EP } (\bibinfo {year}
  {2016}{\natexlab{a}})},\ \bibinfo {note} {article}\BibitemShut {NoStop}%
\bibitem [{\citenamefont {Claughton}\ and\ \citenamefont
  {Lambert}(1996)}]{Claughton1996}%
  \BibitemOpen
  \bibfield  {author} {\bibinfo {author} {\bibfnamefont {N.~R.}\ \bibnamefont
  {Claughton}}\ and\ \bibinfo {author} {\bibfnamefont {C.~J.}\ \bibnamefont
  {Lambert}},\ }\bibfield  {title} {\bibinfo {title} {Thermoelectric properties
  of mesoscopic superconductors},\ }\href
  {https://doi.org/10.1103/PhysRevB.53.6605} {\bibfield  {journal} {\bibinfo
  {journal} {Phys. Rev. B}\ }\textbf {\bibinfo {volume} {53}},\ \bibinfo
  {pages} {6605} (\bibinfo {year} {1996})}\BibitemShut {NoStop}%
\bibitem [{\citenamefont {Anantram}\ and\ \citenamefont
  {Datta}(1996)}]{Anantram_1996}%
  \BibitemOpen
  \bibfield  {author} {\bibinfo {author} {\bibfnamefont {M.~P.}\ \bibnamefont
  {Anantram}}\ and\ \bibinfo {author} {\bibfnamefont {S.}~\bibnamefont
  {Datta}},\ }\bibfield  {title} {\bibinfo {title} {Current fluctuations in
  mesoscopic systems with andreev scattering},\ }\href
  {https://doi.org/10.1103/PhysRevB.53.16390} {\bibfield  {journal} {\bibinfo
  {journal} {Phys. Rev. B}\ }\textbf {\bibinfo {volume} {53}},\ \bibinfo
  {pages} {16390} (\bibinfo {year} {1996})}\BibitemShut {NoStop}%
\bibitem [{\citenamefont {Giazotto}\ \emph {et~al.}(2006)\citenamefont
  {Giazotto}, \citenamefont {Heikkil\"a}, \citenamefont {Luukanen},
  \citenamefont {Savin},\ and\ \citenamefont {Pekola}}]{RevModPhys.78.217}%
  \BibitemOpen
  \bibfield  {author} {\bibinfo {author} {\bibfnamefont {F.}~\bibnamefont
  {Giazotto}}, \bibinfo {author} {\bibfnamefont {T.~T.}\ \bibnamefont
  {Heikkil\"a}}, \bibinfo {author} {\bibfnamefont {A.}~\bibnamefont
  {Luukanen}}, \bibinfo {author} {\bibfnamefont {A.~M.}\ \bibnamefont
  {Savin}},\ and\ \bibinfo {author} {\bibfnamefont {J.~P.}\ \bibnamefont
  {Pekola}},\ }\bibfield  {title} {\bibinfo {title} {Opportunities for
  mesoscopics in thermometry and refrigeration: Physics and applications},\
  }\href {https://doi.org/10.1103/RevModPhys.78.217} {\bibfield  {journal}
  {\bibinfo  {journal} {Rev. Mod. Phys.}\ }\textbf {\bibinfo {volume} {78}},\
  \bibinfo {pages} {217} (\bibinfo {year} {2006})}\BibitemShut {NoStop}%
\bibitem [{\citenamefont {Chandrasekhar}(2009)}]{Chandrasekhar_2009}%
  \BibitemOpen
  \bibfield  {author} {\bibinfo {author} {\bibfnamefont {V.}~\bibnamefont
  {Chandrasekhar}},\ }\bibfield  {title} {\bibinfo {title} {Thermal transport
  in superconductor/normal-metal structures},\ }\href
  {https://doi.org/10.1088/0953-2048/22/8/083001} {\bibfield  {journal}
  {\bibinfo  {journal} {Superconductor Science and Technology}\ }\textbf
  {\bibinfo {volume} {22}},\ \bibinfo {pages} {083001} (\bibinfo {year}
  {2009})}\BibitemShut {NoStop}%
\bibitem [{\citenamefont {Peltonen}\ \emph {et~al.}(2010)\citenamefont
  {Peltonen}, \citenamefont {Virtanen}, \citenamefont {Meschke}, \citenamefont
  {Koski}, \citenamefont {Heikkil\"a},\ and\ \citenamefont
  {Pekola}}]{Peltonen2010}%
  \BibitemOpen
  \bibfield  {author} {\bibinfo {author} {\bibfnamefont {J.~T.}\ \bibnamefont
  {Peltonen}}, \bibinfo {author} {\bibfnamefont {P.}~\bibnamefont {Virtanen}},
  \bibinfo {author} {\bibfnamefont {M.}~\bibnamefont {Meschke}}, \bibinfo
  {author} {\bibfnamefont {J.~V.}\ \bibnamefont {Koski}}, \bibinfo {author}
  {\bibfnamefont {T.~T.}\ \bibnamefont {Heikkil\"a}},\ and\ \bibinfo {author}
  {\bibfnamefont {J.~P.}\ \bibnamefont {Pekola}},\ }\bibfield  {title}
  {\bibinfo {title} {Thermal conductance by the inverse proximity effect in a
  superconductor},\ }\href {https://doi.org/10.1103/PhysRevLett.105.097004}
  {\bibfield  {journal} {\bibinfo  {journal} {Phys. Rev. Lett.}\ }\textbf
  {\bibinfo {volume} {105}},\ \bibinfo {pages} {097004} (\bibinfo {year}
  {2010})}\BibitemShut {NoStop}%
\bibitem [{\citenamefont {Bagrets}\ \emph {et~al.}(2016)\citenamefont
  {Bagrets}, \citenamefont {Altland},\ and\ \citenamefont
  {Kamenev}}]{PhysRevLett.117.196801}%
  \BibitemOpen
  \bibfield  {author} {\bibinfo {author} {\bibfnamefont {D.}~\bibnamefont
  {Bagrets}}, \bibinfo {author} {\bibfnamefont {A.}~\bibnamefont {Altland}},\
  and\ \bibinfo {author} {\bibfnamefont {A.}~\bibnamefont {Kamenev}},\
  }\bibfield  {title} {\bibinfo {title} {Sinai diffusion at quasi-1d
  topological phase transitions},\ }\href
  {https://doi.org/10.1103/PhysRevLett.117.196801} {\bibfield  {journal}
  {\bibinfo  {journal} {Phys. Rev. Lett.}\ }\textbf {\bibinfo {volume} {117}},\
  \bibinfo {pages} {196801} (\bibinfo {year} {2016})}\BibitemShut {NoStop}%
\bibitem [{\citenamefont {Tikhonov}\ \emph {et~al.}(2020)\citenamefont
  {Tikhonov}, \citenamefont {Denisov}, \citenamefont {Piatrusha}, \citenamefont
  {Khrapach}, \citenamefont {Pekola}, \citenamefont {Karimi}, \citenamefont
  {Jabdaraghi},\ and\ \citenamefont {Khrapai}}]{Tikhonov_2020}%
  \BibitemOpen
  \bibfield  {author} {\bibinfo {author} {\bibfnamefont {E.~S.}\ \bibnamefont
  {Tikhonov}}, \bibinfo {author} {\bibfnamefont {A.~O.}\ \bibnamefont
  {Denisov}}, \bibinfo {author} {\bibfnamefont {S.~U.}\ \bibnamefont
  {Piatrusha}}, \bibinfo {author} {\bibfnamefont {I.~N.}\ \bibnamefont
  {Khrapach}}, \bibinfo {author} {\bibfnamefont {J.~P.}\ \bibnamefont
  {Pekola}}, \bibinfo {author} {\bibfnamefont {B.}~\bibnamefont {Karimi}},
  \bibinfo {author} {\bibfnamefont {R.~N.}\ \bibnamefont {Jabdaraghi}},\ and\
  \bibinfo {author} {\bibfnamefont {V.~S.}\ \bibnamefont {Khrapai}},\
  }\bibfield  {title} {\bibinfo {title} {Spatial and energy resolution of
  electronic states by shot noise},\ }\href
  {https://doi.org/10.1103/PhysRevB.102.085417} {\bibfield  {journal} {\bibinfo
   {journal} {Phys. Rev. B}\ }\textbf {\bibinfo {volume} {102}},\ \bibinfo
  {pages} {085417} (\bibinfo {year} {2020})}\BibitemShut {NoStop}%
\bibitem [{\citenamefont {Larocque}\ \emph {et~al.}(2020)\citenamefont
  {Larocque}, \citenamefont {Pinsolle}, \citenamefont {Lupien},\ and\
  \citenamefont {Reulet}}]{Larocque_2020}%
  \BibitemOpen
  \bibfield  {author} {\bibinfo {author} {\bibfnamefont {S.}~\bibnamefont
  {Larocque}}, \bibinfo {author} {\bibfnamefont {E.}~\bibnamefont {Pinsolle}},
  \bibinfo {author} {\bibfnamefont {C.}~\bibnamefont {Lupien}},\ and\ \bibinfo
  {author} {\bibfnamefont {B.}~\bibnamefont {Reulet}},\ }\bibfield  {title}
  {\bibinfo {title} {Shot noise of a temperature-biased tunnel junction},\
  }\href {https://doi.org/10.1103/PhysRevLett.125.106801} {\bibfield  {journal}
  {\bibinfo  {journal} {Phys. Rev. Lett.}\ }\textbf {\bibinfo {volume} {125}},\
  \bibinfo {pages} {106801} (\bibinfo {year} {2020})}\BibitemShut {NoStop}%
\bibitem [{\citenamefont {Landauer}\ and\ \citenamefont
  {Martin}(1991)}]{Landauer_1991}%
  \BibitemOpen
  \bibfield  {author} {\bibinfo {author} {\bibfnamefont {R.}~\bibnamefont
  {Landauer}}\ and\ \bibinfo {author} {\bibfnamefont {T.}~\bibnamefont
  {Martin}},\ }\bibfield  {title} {\bibinfo {title} {Equilibrium and shot noise
  in mesoscopic systems},\ }\href
  {https://doi.org/10.1016/0921-4526(91)90710-v} {\bibfield  {journal}
  {\bibinfo  {journal} {Physica B: Condensed Matter}\ }\textbf {\bibinfo
  {volume} {175}},\ \bibinfo {pages} {167} (\bibinfo {year}
  {1991})}\BibitemShut {NoStop}%
\bibitem [{\citenamefont {Blanter}\ and\ \citenamefont
  {Büttiker}(2000)}]{BLANTER20001}%
  \BibitemOpen
  \bibfield  {author} {\bibinfo {author} {\bibfnamefont {Y.}~\bibnamefont
  {Blanter}}\ and\ \bibinfo {author} {\bibfnamefont {M.}~\bibnamefont
  {Büttiker}},\ }\bibfield  {title} {\bibinfo {title} {Shot noise in
  mesoscopic conductors},\ }\href
  {https://doi.org/https://doi.org/10.1016/S0370-1573(99)00123-4} {\bibfield
  {journal} {\bibinfo  {journal} {Physics Reports}\ }\textbf {\bibinfo {volume}
  {336}},\ \bibinfo {pages} {1 } (\bibinfo {year} {2000})}\BibitemShut
  {NoStop}%
\bibitem [{\citenamefont {Hofstetter}\ \emph {et~al.}(2009)\citenamefont
  {Hofstetter}, \citenamefont {Csonka}, \citenamefont {Nyg{\aa}rd},\ and\
  \citenamefont {Schönenberger}}]{Hofstetter_2009}%
  \BibitemOpen
  \bibfield  {author} {\bibinfo {author} {\bibfnamefont {L.}~\bibnamefont
  {Hofstetter}}, \bibinfo {author} {\bibfnamefont {S.}~\bibnamefont {Csonka}},
  \bibinfo {author} {\bibfnamefont {J.}~\bibnamefont {Nyg{\aa}rd}},\ and\
  \bibinfo {author} {\bibfnamefont {C.}~\bibnamefont {Schönenberger}},\
  }\bibfield  {title} {\bibinfo {title} {Cooper pair splitter realized in a
  two-quantum-dot y-junction},\ }\href {https://doi.org/10.1038/nature08432}
  {\bibfield  {journal} {\bibinfo  {journal} {Nature}\ }\textbf {\bibinfo
  {volume} {461}},\ \bibinfo {pages} {960} (\bibinfo {year}
  {2009})}\BibitemShut {NoStop}%
\bibitem [{\citenamefont {Das}\ \emph {et~al.}(2012{\natexlab{b}})\citenamefont
  {Das}, \citenamefont {Ronen}, \citenamefont {Heiblum}, \citenamefont
  {Mahalu}, \citenamefont {Kretinin},\ and\ \citenamefont
  {Shtrikman}}]{DasDas2012}%
  \BibitemOpen
  \bibfield  {author} {\bibinfo {author} {\bibfnamefont {A.}~\bibnamefont
  {Das}}, \bibinfo {author} {\bibfnamefont {Y.}~\bibnamefont {Ronen}}, \bibinfo
  {author} {\bibfnamefont {M.}~\bibnamefont {Heiblum}}, \bibinfo {author}
  {\bibfnamefont {D.}~\bibnamefont {Mahalu}}, \bibinfo {author} {\bibfnamefont
  {A.~V.}\ \bibnamefont {Kretinin}},\ and\ \bibinfo {author} {\bibfnamefont
  {H.}~\bibnamefont {Shtrikman}},\ }\bibfield  {title} {\bibinfo {title}
  {High-efficiency cooper pair splitting demonstrated by two-particle
  conductance resonance and positive noise cross-correlation},\ }\href
  {https://doi.org/10.1038/ncomms2169} {\bibfield  {journal} {\bibinfo
  {journal} {Nature Communications}\ }\textbf {\bibinfo {volume} {3}},\
  \bibinfo {pages} {1165 EP } (\bibinfo {year} {2012}{\natexlab{b}})},\
  \bibinfo {note} {article}\BibitemShut {NoStop}%
\bibitem [{\citenamefont {Deng}\ \emph {et~al.}(2016)\citenamefont {Deng},
  \citenamefont {Vaitiekenas}, \citenamefont {Hansen}, \citenamefont {Danon},
  \citenamefont {Leijnse}, \citenamefont {Flensberg}, \citenamefont {Nyg{\r
  a}rd}, \citenamefont {Krogstrup},\ and\ \citenamefont {Marcus}}]{Deng1557}%
  \BibitemOpen
  \bibfield  {author} {\bibinfo {author} {\bibfnamefont {M.~T.}\ \bibnamefont
  {Deng}}, \bibinfo {author} {\bibfnamefont {S.}~\bibnamefont {Vaitiekenas}},
  \bibinfo {author} {\bibfnamefont {E.~B.}\ \bibnamefont {Hansen}}, \bibinfo
  {author} {\bibfnamefont {J.}~\bibnamefont {Danon}}, \bibinfo {author}
  {\bibfnamefont {M.}~\bibnamefont {Leijnse}}, \bibinfo {author} {\bibfnamefont
  {K.}~\bibnamefont {Flensberg}}, \bibinfo {author} {\bibfnamefont
  {J.}~\bibnamefont {Nyg{\r a}rd}}, \bibinfo {author} {\bibfnamefont
  {P.}~\bibnamefont {Krogstrup}},\ and\ \bibinfo {author} {\bibfnamefont
  {C.~M.}\ \bibnamefont {Marcus}},\ }\bibfield  {title} {\bibinfo {title}
  {Majorana bound state in a coupled quantum-dot hybrid-nanowire system},\
  }\href {https://doi.org/10.1126/science.aaf3961} {\bibfield  {journal}
  {\bibinfo  {journal} {Science}\ }\textbf {\bibinfo {volume} {354}},\ \bibinfo
  {pages} {1557} (\bibinfo {year} {2016})}\BibitemShut {NoStop}%
\bibitem [{\citenamefont {Albrecht}\ \emph {et~al.}(2016)\citenamefont
  {Albrecht}, \citenamefont {Higginbotham}, \citenamefont {Madsen},
  \citenamefont {Kuemmeth}, \citenamefont {Jespersen}, \citenamefont
  {Nyg{\aa}rd}, \citenamefont {Krogstrup},\ and\ \citenamefont
  {Marcus}}]{Albrecht2016}%
  \BibitemOpen
  \bibfield  {author} {\bibinfo {author} {\bibfnamefont {S.~M.}\ \bibnamefont
  {Albrecht}}, \bibinfo {author} {\bibfnamefont {A.~P.}\ \bibnamefont
  {Higginbotham}}, \bibinfo {author} {\bibfnamefont {M.}~\bibnamefont
  {Madsen}}, \bibinfo {author} {\bibfnamefont {F.}~\bibnamefont {Kuemmeth}},
  \bibinfo {author} {\bibfnamefont {T.~S.}\ \bibnamefont {Jespersen}}, \bibinfo
  {author} {\bibfnamefont {J.}~\bibnamefont {Nyg{\aa}rd}}, \bibinfo {author}
  {\bibfnamefont {P.}~\bibnamefont {Krogstrup}},\ and\ \bibinfo {author}
  {\bibfnamefont {C.~M.}\ \bibnamefont {Marcus}},\ }\bibfield  {title}
  {\bibinfo {title} {Exponential protection of zero modes in majorana
  islands},\ }\href {https://doi.org/10.1038/nature17162} {\bibfield  {journal}
  {\bibinfo  {journal} {Nature}\ }\textbf {\bibinfo {volume} {531}},\ \bibinfo
  {pages} {206} (\bibinfo {year} {2016})}\BibitemShut {NoStop}%
\bibitem [{\citenamefont {Denisov}\ \emph {et~al.}(2020)\citenamefont
  {Denisov}, \citenamefont {Bubis}, \citenamefont {Piatrusha}, \citenamefont
  {Titova}, \citenamefont {Nasibulin}, \citenamefont {Becker}, \citenamefont
  {Treu}, \citenamefont {Ruhstorfer}, \citenamefont {Koblmueller},
  \citenamefont {Tikhonov},\ and\ \citenamefont {Khrapai}}]{Denisov2020}%
  \BibitemOpen
  \bibfield  {author} {\bibinfo {author} {\bibfnamefont {A.~O.}\ \bibnamefont
  {Denisov}}, \bibinfo {author} {\bibfnamefont {A.~V.}\ \bibnamefont {Bubis}},
  \bibinfo {author} {\bibfnamefont {S.~U.}\ \bibnamefont {Piatrusha}}, \bibinfo
  {author} {\bibfnamefont {N.~A.}\ \bibnamefont {Titova}}, \bibinfo {author}
  {\bibfnamefont {A.~G.}\ \bibnamefont {Nasibulin}}, \bibinfo {author}
  {\bibfnamefont {J.}~\bibnamefont {Becker}}, \bibinfo {author} {\bibfnamefont
  {J.}~\bibnamefont {Treu}}, \bibinfo {author} {\bibfnamefont {D.}~\bibnamefont
  {Ruhstorfer}}, \bibinfo {author} {\bibfnamefont {G.}~\bibnamefont
  {Koblmueller}}, \bibinfo {author} {\bibfnamefont {E.~S.}\ \bibnamefont
  {Tikhonov}},\ and\ \bibinfo {author} {\bibfnamefont {V.~S.}\ \bibnamefont
  {Khrapai}},\ }\bibfield  {title} {\bibinfo {title} {Heat-mode excitation in a
  proximity superconductor},\ }\Eprint {https://arxiv.org/abs/2006.09803}
  {arXiv:2006.09803}  (\bibinfo {year} {2020})\BibitemShut {NoStop}%
\bibitem [{\citenamefont {Ford}\ \emph {et~al.}(2009)\citenamefont {Ford},
  \citenamefont {Ho}, \citenamefont {Chueh}, \citenamefont {Tseng},
  \citenamefont {Fan}, \citenamefont {Guo}, \citenamefont {Bokor},\ and\
  \citenamefont {Javey}}]{Ford_2009}%
  \BibitemOpen
  \bibfield  {author} {\bibinfo {author} {\bibfnamefont {A.~C.}\ \bibnamefont
  {Ford}}, \bibinfo {author} {\bibfnamefont {J.~C.}\ \bibnamefont {Ho}},
  \bibinfo {author} {\bibfnamefont {Y.-L.}\ \bibnamefont {Chueh}}, \bibinfo
  {author} {\bibfnamefont {Y.-C.}\ \bibnamefont {Tseng}}, \bibinfo {author}
  {\bibfnamefont {Z.}~\bibnamefont {Fan}}, \bibinfo {author} {\bibfnamefont
  {J.}~\bibnamefont {Guo}}, \bibinfo {author} {\bibfnamefont {J.}~\bibnamefont
  {Bokor}},\ and\ \bibinfo {author} {\bibfnamefont {A.}~\bibnamefont {Javey}},\
  }\bibfield  {title} {\bibinfo {title} {Diameter-dependent electron mobility
  of {InAs} nanowires},\ }\href {https://doi.org/10.1021/nl803154m} {\bibfield
  {journal} {\bibinfo  {journal} {Nano Letters}\ }\textbf {\bibinfo {volume}
  {9}},\ \bibinfo {pages} {360} (\bibinfo {year} {2009})}\BibitemShut {NoStop}%
\bibitem [{\citenamefont {Lachenmann}\ \emph {et~al.}(1997)\citenamefont
  {Lachenmann}, \citenamefont {Friedrich}, \citenamefont {F\"orster},
  \citenamefont {Uhlisch},\ and\ \citenamefont {Golubov}}]{PhysRevB.56.14108}%
  \BibitemOpen
  \bibfield  {author} {\bibinfo {author} {\bibfnamefont {S.~G.}\ \bibnamefont
  {Lachenmann}}, \bibinfo {author} {\bibfnamefont {I.}~\bibnamefont
  {Friedrich}}, \bibinfo {author} {\bibfnamefont {A.}~\bibnamefont
  {F\"orster}}, \bibinfo {author} {\bibfnamefont {D.}~\bibnamefont {Uhlisch}},\
  and\ \bibinfo {author} {\bibfnamefont {A.~A.}\ \bibnamefont {Golubov}},\
  }\bibfield  {title} {\bibinfo {title} {Charge transport in
  superconductor/semiconductor/normal-conductor step junctions},\ }\href
  {https://doi.org/10.1103/PhysRevB.56.14108} {\bibfield  {journal} {\bibinfo
  {journal} {Phys. Rev. B}\ }\textbf {\bibinfo {volume} {56}},\ \bibinfo
  {pages} {14108} (\bibinfo {year} {1997})}\BibitemShut {NoStop}%
\bibitem [{\citenamefont {Courtois}\ \emph {et~al.}(1999)\citenamefont
  {Courtois}, \citenamefont {Charlat}, \citenamefont {Gandit}, \citenamefont
  {Mailly},\ and\ \citenamefont {Pannetier}}]{Courtois1999}%
  \BibitemOpen
  \bibfield  {author} {\bibinfo {author} {\bibfnamefont {H.}~\bibnamefont
  {Courtois}}, \bibinfo {author} {\bibfnamefont {P.}~\bibnamefont {Charlat}},
  \bibinfo {author} {\bibfnamefont {P.}~\bibnamefont {Gandit}}, \bibinfo
  {author} {\bibfnamefont {D.}~\bibnamefont {Mailly}},\ and\ \bibinfo {author}
  {\bibfnamefont {B.}~\bibnamefont {Pannetier}},\ }\href
  {https://doi.org/10.1023/a:1021885617107} {\bibfield  {journal} {\bibinfo
  {journal} {Journal of Low Temperature Physics}\ }\textbf {\bibinfo {volume}
  {116}},\ \bibinfo {pages} {187} (\bibinfo {year} {1999})}\BibitemShut
  {NoStop}%
\bibitem [{\citenamefont {Beenakker}(1997)}]{Beenakker_1997}%
  \BibitemOpen
  \bibfield  {author} {\bibinfo {author} {\bibfnamefont {C.~W.~J.}\
  \bibnamefont {Beenakker}},\ }\bibfield  {title} {\bibinfo {title}
  {Random-matrix theory of quantum transport},\ }\href
  {https://doi.org/10.1103/revmodphys.69.731} {\bibfield  {journal} {\bibinfo
  {journal} {Reviews of Modern Physics}\ }\textbf {\bibinfo {volume} {69}},\
  \bibinfo {pages} {731} (\bibinfo {year} {1997})}\BibitemShut {NoStop}%
\bibitem [{\citenamefont {Jehl}\ \emph {et~al.}(2000)\citenamefont {Jehl},
  \citenamefont {Sanquer}, \citenamefont {Calemczuk},\ and\ \citenamefont
  {Mailly}}]{Jehl2000}%
  \BibitemOpen
  \bibfield  {author} {\bibinfo {author} {\bibfnamefont {X.}~\bibnamefont
  {Jehl}}, \bibinfo {author} {\bibfnamefont {M.}~\bibnamefont {Sanquer}},
  \bibinfo {author} {\bibfnamefont {R.}~\bibnamefont {Calemczuk}},\ and\
  \bibinfo {author} {\bibfnamefont {D.}~\bibnamefont {Mailly}},\ }\bibfield
  {title} {\bibinfo {title} {Detection of doubled shot noise in short
  normal-metal/ superconductor junctions},\ }\href
  {https://doi.org/10.1038/35011012} {\bibfield  {journal} {\bibinfo  {journal}
  {Nature}\ }\textbf {\bibinfo {volume} {405}},\ \bibinfo {pages} {50}
  (\bibinfo {year} {2000})}\BibitemShut {NoStop}%
\bibitem [{\citenamefont {Kozhevnikov}\ \emph {et~al.}(2000)\citenamefont
  {Kozhevnikov}, \citenamefont {Schoelkopf},\ and\ \citenamefont
  {Prober}}]{PhysRevLett.84.3398}%
  \BibitemOpen
  \bibfield  {author} {\bibinfo {author} {\bibfnamefont {A.~A.}\ \bibnamefont
  {Kozhevnikov}}, \bibinfo {author} {\bibfnamefont {R.~J.}\ \bibnamefont
  {Schoelkopf}},\ and\ \bibinfo {author} {\bibfnamefont {D.~E.}\ \bibnamefont
  {Prober}},\ }\bibfield  {title} {\bibinfo {title} {Observation of
  photon-assisted noise in a diffusive normal metal--superconductor junction},\
  }\href {https://doi.org/10.1103/PhysRevLett.84.3398} {\bibfield  {journal}
  {\bibinfo  {journal} {Phys. Rev. Lett.}\ }\textbf {\bibinfo {volume} {84}},\
  \bibinfo {pages} {3398} (\bibinfo {year} {2000})}\BibitemShut {NoStop}%
\bibitem [{\citenamefont {Choi}\ \emph {et~al.}(2005)\citenamefont {Choi},
  \citenamefont {Hansen}, \citenamefont {Kontos}, \citenamefont {Hoffmann},
  \citenamefont {Oberholzer}, \citenamefont {Belzig}, \citenamefont
  {Sch\"onenberger}, \citenamefont {Akazaki},\ and\ \citenamefont
  {Takayanagi}}]{PhysRevB.72.024501}%
  \BibitemOpen
  \bibfield  {author} {\bibinfo {author} {\bibfnamefont {B.-R.}\ \bibnamefont
  {Choi}}, \bibinfo {author} {\bibfnamefont {A.~E.}\ \bibnamefont {Hansen}},
  \bibinfo {author} {\bibfnamefont {T.}~\bibnamefont {Kontos}}, \bibinfo
  {author} {\bibfnamefont {C.}~\bibnamefont {Hoffmann}}, \bibinfo {author}
  {\bibfnamefont {S.}~\bibnamefont {Oberholzer}}, \bibinfo {author}
  {\bibfnamefont {W.}~\bibnamefont {Belzig}}, \bibinfo {author} {\bibfnamefont
  {C.}~\bibnamefont {Sch\"onenberger}}, \bibinfo {author} {\bibfnamefont
  {T.}~\bibnamefont {Akazaki}},\ and\ \bibinfo {author} {\bibfnamefont
  {H.}~\bibnamefont {Takayanagi}},\ }\bibfield  {title} {\bibinfo {title}
  {Shot-noise and conductance measurements of transparent
  superconductor/two-dimensional electron gas junctions},\ }\href
  {https://doi.org/10.1103/PhysRevB.72.024501} {\bibfield  {journal} {\bibinfo
  {journal} {Phys. Rev. B}\ }\textbf {\bibinfo {volume} {72}},\ \bibinfo
  {pages} {024501} (\bibinfo {year} {2005})}\BibitemShut {NoStop}%
\bibitem [{\citenamefont {Tikhonov}\ \emph
  {et~al.}(2016{\natexlab{b}})\citenamefont {Tikhonov}, \citenamefont
  {Shovkun}, \citenamefont {Snelder}, \citenamefont {Stehno}, \citenamefont
  {Huang}, \citenamefont {Golden}, \citenamefont {Golubov}, \citenamefont
  {Brinkman},\ and\ \citenamefont {Khrapai}}]{Tikhonov_PRL_2016}%
  \BibitemOpen
  \bibfield  {author} {\bibinfo {author} {\bibfnamefont {E.~S.}\ \bibnamefont
  {Tikhonov}}, \bibinfo {author} {\bibfnamefont {D.~V.}\ \bibnamefont
  {Shovkun}}, \bibinfo {author} {\bibfnamefont {M.}~\bibnamefont {Snelder}},
  \bibinfo {author} {\bibfnamefont {M.~P.}\ \bibnamefont {Stehno}}, \bibinfo
  {author} {\bibfnamefont {Y.}~\bibnamefont {Huang}}, \bibinfo {author}
  {\bibfnamefont {M.~S.}\ \bibnamefont {Golden}}, \bibinfo {author}
  {\bibfnamefont {A.~A.}\ \bibnamefont {Golubov}}, \bibinfo {author}
  {\bibfnamefont {A.}~\bibnamefont {Brinkman}},\ and\ \bibinfo {author}
  {\bibfnamefont {V.~S.}\ \bibnamefont {Khrapai}},\ }\bibfield  {title}
  {\bibinfo {title} {Andreev reflection in an $s$-type superconductor
  proximized 3d topological insulator},\ }\href
  {https://doi.org/10.1103/PhysRevLett.117.147001} {\bibfield  {journal}
  {\bibinfo  {journal} {Phys. Rev. Lett.}\ }\textbf {\bibinfo {volume} {117}},\
  \bibinfo {pages} {147001} (\bibinfo {year} {2016}{\natexlab{b}})}\BibitemShut
  {NoStop}%
\bibitem [{\citenamefont {Haim}\ and\ \citenamefont {Stern}(2019)}]{Haim_2019}%
  \BibitemOpen
  \bibfield  {author} {\bibinfo {author} {\bibfnamefont {A.}~\bibnamefont
  {Haim}}\ and\ \bibinfo {author} {\bibfnamefont {A.}~\bibnamefont {Stern}},\
  }\bibfield  {title} {\bibinfo {title} {Benefits of weak disorder in
  one-dimensional topological superconductors},\ }\href
  {https://doi.org/10.1103/PhysRevLett.122.126801} {\bibfield  {journal}
  {\bibinfo  {journal} {Phys. Rev. Lett.}\ }\textbf {\bibinfo {volume} {122}},\
  \bibinfo {pages} {126801} (\bibinfo {year} {2019})}\BibitemShut {NoStop}%
\bibitem [{\citenamefont {Laeven}\ \emph {et~al.}(2020)\citenamefont {Laeven},
  \citenamefont {Nijholt}, \citenamefont {Wimmer},\ and\ \citenamefont
  {Akhmerov}}]{Laeven_2020}%
  \BibitemOpen
  \bibfield  {author} {\bibinfo {author} {\bibfnamefont {T.}~\bibnamefont
  {Laeven}}, \bibinfo {author} {\bibfnamefont {B.}~\bibnamefont {Nijholt}},
  \bibinfo {author} {\bibfnamefont {M.}~\bibnamefont {Wimmer}},\ and\ \bibinfo
  {author} {\bibfnamefont {A.~R.}\ \bibnamefont {Akhmerov}},\ }\bibfield
  {title} {\bibinfo {title} {Enhanced proximity effect in zigzag-shaped
  majorana josephson junctions},\ }\href
  {https://doi.org/10.1103/PhysRevLett.125.086802} {\bibfield  {journal}
  {\bibinfo  {journal} {Phys. Rev. Lett.}\ }\textbf {\bibinfo {volume} {125}},\
  \bibinfo {pages} {086802} (\bibinfo {year} {2020})}\BibitemShut {NoStop}%
\bibitem [{\citenamefont {Franz}\ and\ \citenamefont
  {Wiedemann}(1853)}]{Franz_1853}%
  \BibitemOpen
  \bibfield  {author} {\bibinfo {author} {\bibfnamefont {R.}~\bibnamefont
  {Franz}}\ and\ \bibinfo {author} {\bibfnamefont {G.}~\bibnamefont
  {Wiedemann}},\ }\bibfield  {title} {\bibinfo {title} {Ueber die
  wärme-leitungsfähigkeit der metalle},\ }\href
  {https://doi.org/10.1002/andp.18531650802} {\bibfield  {journal} {\bibinfo
  {journal} {Annalen der Physik und Chemie}\ }\textbf {\bibinfo {volume}
  {165}},\ \bibinfo {pages} {497} (\bibinfo {year} {1853})}\BibitemShut
  {NoStop}%
\end{thebibliography}

%

\end{document}


\widetext
	\clearpage
	\begin{center}
		\textbf{\large Supplemental Material}
	\end{center}
	\setcounter{equation}{0}
	\setcounter{figure}{0}
	\setcounter{table}{0}
	\setcounter{page}{1}
	\makeatletter
	\renewcommand{\theequation}{S\arabic{equation}}
	\renewcommand{\figurename}{Supplemental Material Fig.}
	\renewcommand{\bibnumfmt}[1]{[S#1]}
	\renewcommand{\citenumfont}[1]{S#1}
	
	\newcommand{\mysub}[2]{#1_{\text{#2}}}
	\newcommand{\myvareq}[4]{$#1_{\text{#2}}=#3\,\mathrm{#4}$}
	\newcommand{\myq}[2]{$#1\,\mathrm{#2}$}
	\newcommand{\ED}[3]{f^{#1}_{#2,#3}}
	\section*{Differential resistance and conductance matrices.}

Here we outline a relation between the quantities measured in transport experiments with either fixed currents or fixed voltages applied to the normal terminals 1 and 2 in the NSN device. The voltages $V_1,\, V_2$ are measured with respect to the (grounded) S terminal. The positive direction of the currents $I_1,\, I_2$ is from the respective lead into the device, ensuring positive signs of the diagonal transport responses. 

In a conductance experiment the independent (externally controlled) variables are voltages and the currents are measured. The elements of the differential conductance matrix are:

\begin{align*}
	G_{11} \equiv & \left.\frac{\partial I_1}{\partial V_1}\right|_{V_2=const} & G_{12} \equiv  \left.\frac{\partial I_1}{\partial V_2}\right|_{V_1=const} \\
	G_{21} \equiv & \left.\frac{\partial I_2}{\partial V_1}\right|_{V_2=const}	& G_{22} \equiv \left.\frac{\partial I_2}{\partial V_2}\right|_{V_1=const}
\end{align*}

A relation to the resistance experiment is straightforward. In this case externally controlled are currents and measured are voltages. As follows from the conductance matrix, the condition $I_2=const$ is satisfied provided  $G_{21}\delta V_1 +G_{22}\delta V_2 \equiv \delta I_2 = 0$. With this constraint we have:

\begin{equation*}
 \delta I_1 = G_{11}\delta V_1 + G_{12}\delta V_2 = \frac{\mathrm{DET}}{G_{22}}\delta V_1, 
\end{equation*}
where $\delta$ denotes an infinitesimal increment of the current/voltage and  DET is the determinant of the conductance matrix $\mathrm{DET} \equiv G_{11}G_{22} - G_{12}G_{21}$. Similar relation holds for $\delta I_2$ under condition of $I_1=const$. Thus, for the differential resistance matrix we obtain:

\begin{align*}
	R_{11} \equiv & \left.\frac{\partial V_1}{\partial I_1}\right|_{I_2=const} = \frac{G_{22}}{\mathrm{DET}} &R_{12} \equiv & \left.\frac{\partial V_1}{\partial I_2}\right|_{I_1=const}= -\frac{G_{12}}{\mathrm{DET}} \\
	R_{21} \equiv & \left.\frac{\partial V_2}{\partial I_1}\right|_{I_2=const} = -\frac{G_{21}}{\mathrm{DET}} &R_{22} \equiv & \left.\frac{\partial V_2}{\partial I_2}\right|_{I_1=const}=\frac{G_{11}}{\mathrm{DET}}
\end{align*} 

In present experiment the experimentally measured non-local (non-diagonal) diferential resistances are always much smaller compared to the diagonal ones. It follows that $|G_{12}|,|G_{21}| \ll |G_{22}|,|G_{11}|$, so that $\mathrm{DET}\approx G_{11}G_{22}$. This allows to obtain the relations from the main manuscript $G_{ii}\approx R_{ii}^{-1}$ and $G_{ij} \approx -R_{ij}/(R_{11}R_{22}),\, i\neq j$.

	\section*{Nonlocal I-Vs and differential conductance.}
	
	\begin{figure*}[h]
	\begin{center}
		\includegraphics[scale=1]{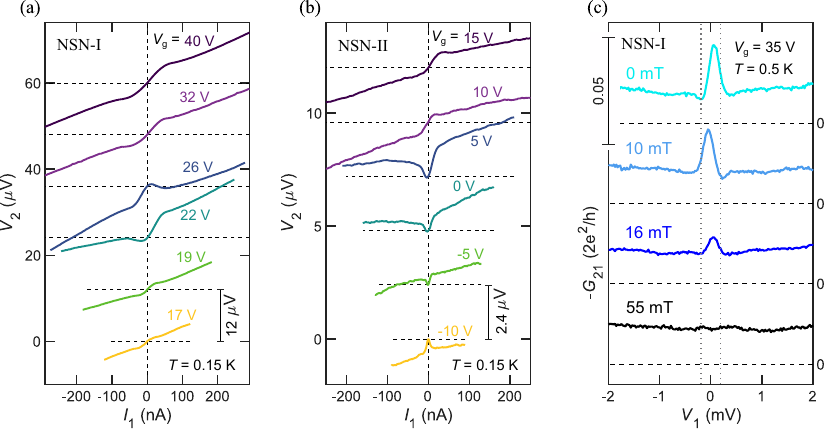}
	\end{center}
	\caption{{(a, b)Measured nonlocal I-V characteristics versus $V_{\mathrm{g}}$ for both devices in zero magnetic field. Different curves are vertically spaced with zero level shown as the dashed lines.~(c) Non-diagonal conductance is plotted as a function of the bias voltage at different magnetic fields. Dotted lines show position of the superconducting gap.}}
	\label{sup_fig1}
\end{figure*}
Measured $V_{2}(I_{1})$ curves in the nonlocal configurations show similar non-universal behavior for both devices as evident from Fig.~\ref{sup_fig1}a, b. First, the nonlocal voltage signal is small $|V_{2}|<k_{\mathrm{B}}T/e\approx 13~\mu V$ and we can suppose the junction N$_2$-NW-S to be essentially unbiased. Second, some of the I-V curves are highly non-anti-symmetric as a results we observe finite odd contribution to the non-diagonal conductance $G_{12}$ as shown in Fig.~\ref{sup_fig1}c. At increasing magnetic field all the sub-gap features are smeared and eventually remain featureless in sufficiently high $B~=~55~\mathrm{mT}$, where the superconductivity of Al is already suppressed.
\section{\textcolor{black}{Device fabrication}}
{InAs nanowires grown by molecular beam epitaxy on Si substrate~\cite{Hertenberger_2010} are ultrasonicated in isopropyl alcohol. Nanowires are drop casted on Si/SiO2 (300 nm) substrates~\cite{PhysRevB.97.115306} with preliminary defined alignment marks.} For superconducting contacts conventional electron beam lithography (EBL) followed by e-beam deposition of Al (150\,nm) is utilized. {To} obtain {the} ohmic contacts{,} in-situ Ar ion milling is performed before Al deposition in a chamber with a base pressure below $10^{-7}$\,mbar. Normal metal contacts are fabricated in two different ways (different device batches): magnetron sputtering or e-beam deposition. For sputtering (NS and NSN~-~I devices) in-situ Ar plasma etching is followed by sputtering of Ti/Au (5\,nm/200\,nm). Normal metal contacts Ti/Au (5\,nm/150\,nm) in device NSN~-~II are deposited in the same way {as} superconducting ones.
	\newpage
	\clearpage
	\section*{Current transfer length estimation}
	\begin{figure*}[h]
		\begin{center}
			\includegraphics[scale=1]{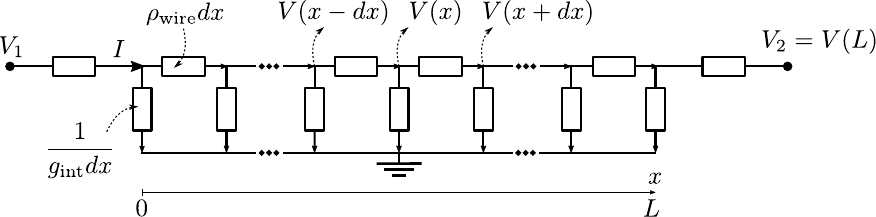}
		\end{center}
		\caption{\textbf{Effective resistance model for nanowire/superconductor interface.} }
		\label{sup_fig1}
	\end{figure*}
	
	To estimate the characteristic length of charge overflow within grounded S terminal $l_{\mathrm{T}}$ we use circuit shown in fig.\,\ref{sup_fig1}. Here $\rho_{\mathrm{wire}}$ and $g_{\mathrm{int}}$ are resistance of the nanowire (NW) and conductivity of interface per unit length respectively. In the continuous limit we can write current conservation for each point along NW/S interface:
	\begin{equation*}
		\begin{gathered}
			\frac{V(x+dx)-V(x)}{\rho_{\mathrm{wire}} dx}+\frac{V(x-dx)-V(x)}{\rho_{\mathrm{wire}} dx}=\frac{V(x)dx}{1/g_{\mathrm{int}}} \\ 
			\frac{d^2V(x)}{dx^2}=\frac{V(x)}{l_{\mathrm{T}}^2},~~l_{\mathrm{T}}=\frac{1}{\sqrt{\rho_{\mathrm{wire}}g_{\mathrm{int}}}}
		\end{gathered}
	\end{equation*}
	Boundary conditions including one that normal terminal N2 is floating and no current flow into it. \begin{equation*}
		\frac{dV(x)}{dx}\bigg{|}_{x=0}=-\rho_{\mathrm{wire}} I,\;\; 
		\frac{dV(x)}{dx}\bigg{|}_{x=L}=0 
	\end{equation*}
	Solving elementary Neumann problem we can find non-local rsistance $r_{21}$:
	\begin{equation*}
		\begin{gathered}
			R_{21}=\frac{V(L)}{I}=\frac{l_{\mathrm{T}}\rho_{\mathrm{wire}}}{\sinh(\frac{L}{l_{\mathrm{T}}})}
		\end{gathered}
	\end{equation*}
	At high biases $|V_1|\gg\Delta/e$ for two measured devices (NSN-I, NSN-II) we have $R_{21}\approx40,~10\,\Omega$ and $L\approx200,~300\,\mathrm{nm}$ respectively, thus $l_{\mathrm{T}}\approx75\,\mathrm{nm}$.

		\newpage
		
	\section{\textcolor{black}{Critical Temperature of Al contacts}}
	\begin{figure*}[h]
		\begin{center}
			\vspace{-0.6cm}
			\includegraphics[scale=1]{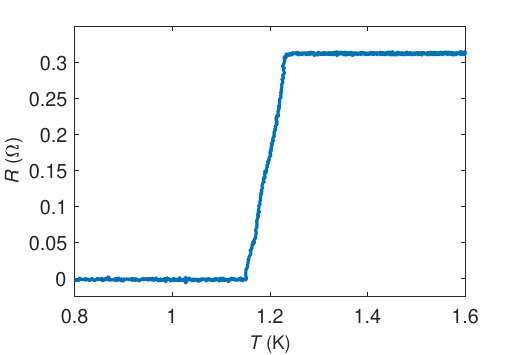}
			\vspace{-0.4cm}
			\caption{The resistance of a four-terminal Al strip, deposited via the same process, as the one used in the fabrication of the samples, featured in the main text.}
			\label{sup_fig_al}
		\end{center}
	\end{figure*}
	
	The temperature dependence of superconducting Al, deposited via the same process as described in "Device Fabrication" was performed separately on the four-terminal Al strips, incorporated in the samples studied in~\cite{Bubis_2017}. Here we present raw data (see Fig.~\ref{sup_fig_al}), which leads to the estimate $T_\mathrm{c}~ =~ 1.20 \pm 0.03$\,K, corresponding to the superconducting energy gap of $\Delta~=~183\pm5~\mu V$ in Al leads.

	
	\renewcommand*{\bibfont}{\small}
	
%